\documentclass[final,5p,times,twocolumn]{elsarticle}
\usepackage{amssymb}
\usepackage{amsmath}
\usepackage{bm}
\def\ve#1{{\bm{#1}}}
\def\nuc#1#2#3{{}^{#2}_{#3}\mathrm{#1}}
\def\urm#1{\scriptstyle{\text{\textrm{\textmd{\textup{#1}}}}}}

\let\Re\relax
\DeclareMathOperator{\Re}{Re}
\usepackage{siunitx}
\usepackage{booktabs}
\usepackage{longtable}
\usepackage{threeparttable}
\usepackage{dcolumn}
\biboptions{sort&compress}
\journal{Atomic and Nuclear Data Tables}
\begin{document}
\begin{frontmatter}
  \title{Comprehensive Table of Calculated Huff Factors}
  \author[aff1]{Yuichi Uesaka}
  \affiliation[aff1]{organization={Department of Fundamental Education, Dokkyo Medical University},
    addressline={Kitakobayashi 880},
    city={Mibu, Shimotsuga},
    postcode={321-0293},
    state={Tochigi},
    country={Japan}}
  \author[aff2,aff6,aff3]{Tomoya Naito}
  \affiliation[aff2]{organization={Department of Nuclear Engineering and Management, Graduate School of Engineering, The University of Tokyo},
    addressline={Hongo 7-3-1},
    city={Bunkyo},
    postcode={113-8656},
    state={Tokyo},
    country={Japan}}
  \affiliation[aff6]{organization={Department of Physics, Graduate School of Science, The University of Tokyo},
    addressline={Hongo 7-3-1},
    city={Bunkyo},
    postcode={113-0033},
    state={Tokyo},
    country={Japan}}
  \affiliation[aff3]{organization={RIKEN Center for Interdisciplinary Theoretical and Mathematical Sciences (iTHEMS)},
    addressline={Hirosawa 2-1},
    city={Wako},
    postcode={351-0198},
    state={Saitama},
    country={Japan}}
  \author[aff4]{Shuichiro Ebata}
  \affiliation[aff4]{organization={Saitama University},
    addressline={Shimo-Okubo 255},
    city={Sakura-ku, Saitama-shi},
    postcode={338-8570},
    state={Saitama},
    country={Japan}}
  \author[aff5]{Megumi Niikura}
  \affiliation[aff5]{organization={Nishina Center, RIKEN},
    addressline={Hirosawa 2-1},
    city={Wako},
    postcode={351-0198},
    state={Saitama},
    country={Japan}}
  % 
  %% Abstract
  \begin{abstract}
    We present a systematic calculation of the Huff factor for nuclei with atomic numbers ($ Z $) in the range of $ 6 \leq Z \leq 94 $.
    The Huff factor quantifies the increase in the partial lifetime of the decay-in-orbit (DIO) of the muonic atom and serves as an essential correction factor for extracting the nuclear muon capture rate from the measured lifetimes of the muonic atom.
    However, previous calculations typically provided only the atomic number dependence and neglected isotope dependence---an assumption whose reliability had not been examined, despite its importance for a comprehensive understanding of the nuclear muon capture rate.
    In this work, we calculate the Huff factor using nuclear charge distributions obtained from a fully self-consistent microscopic nuclear structure model that incorporates pairing and deformation effects.
    The resultant Huff factors exhibit a monotonic decrease with increasing $ Z $, while the isotope dependence is found to be small.
    Our results also show good agreement with previous calculations, supporting the reliability of the present framework.
    The comprehensive set of Huff factors presented here constitutes the first unified values currently available and will serve as a basis for future evaluations of muon nuclear data.
  \end{abstract}
\end{frontmatter}
% 
%% ==================  main text ====================
% 
\section{Introduction}
\label{sec:introduction}
\par
A muonic atom consists of a negative muon orbiting an atomic nucleus.
The muon in the $ 1s $ state undergoes two competing decay processes.
One process is the conversion of the muon in an atomic orbit into an electron and two neutrinos, which is known as decay-in-orbit (DIO),
\begin{equation}
  \mu^{-} \rightarrow e^{-} + \bar{\nu}_e + \nu_{\mu}.
\end{equation}
The other is nuclear muon capture, a process in which the muon and a proton in the atomic nucleus convert into a neutron and a muon neutrino,
\begin{equation}
  \mu^{-} + p \rightarrow n + \nu_{\mu}.
\end{equation}
The muon capture rate ($ \Lambda_{\urm{cap}} $) can be obtained from the experimentally measured lifetime of the muonic atom ($ \tau_{\urm{total}} $) as
\begin{equation}
  \frac{1}{\tau_{\urm{total}}}
  =
  \Lambda_{\urm{cap}}
  +
  \frac{Q}{\tau_{\mu^{+}}},
\end{equation}
where $ \tau_{\mu^{+}} $ is the lifetime of the positive muon, $ 2.1969811(22) \, \mathrm{\mu s} $~\cite{pdg2024},
and $ Q $ is the Huff factor~\cite{Huff1961-ur},
a correction factor that accounts for the increase in the partial lifetime of the DIO process due to the muon's binding to an atomic orbital.
Since theoretical calculations provide $ \Lambda_{\urm{cap}} $ directly,
it is crucial to compare experimental data with theoretical predictions of $ \Lambda_{\urm{cap}} $ rather than $ \tau_{\urm{total}} $.
As $ \tau_{\mu^{+}} $ is known with very high precision,
a reliable value of the Huff factor over a wide range of nuclei is necessary for accurate derivation of $ \Lambda_{\urm{cap}} $ from the experimentally measured $ \tau_{\urm{total}} $.
\par
Experimental and theoretical investigations related to muon capture reactions have advanced in recent years, including studies of the particle emission probability~\cite{Manabe2023-zx,Saito2025-mx,Minato2023Phys.Rev.C107_054314} and the production branching ratios of residual nuclei~\cite{Niikura2024-ec,Yamaguchi2025-wr,Mizuno2025-tg,Maekawa2025-cu}.
These quantities are defined relative to the number of captured muons, and their interpretation therefore relies on the capture branching ratio ($ \tau_{\urm{total}} \Lambda_{\urm{cap}} $).
Consequently, a comprehensive and reliable dataset of the Huff factor with well-quantified accuracy is required.
\par
Comprehensive measurements of $ \tau_{\urm{total}} $ have been conducted in the past,
with most results summarized in Ref.~\cite{Suzuki1987-aq}, and recent data included in Ref.~\cite{Iwamoto2025-cf}.
It should be noted that these compilations do not always account for the Huff factor,
with some studies omitting it by assuming $ Q = 1 $.
This inconsistency in treating the Huff factor leads to the problem that the conversion from
$ \tau_{\urm{total}} $ to $ \Lambda_{\urm{cap}} $ is not performed in a unified manner.
The Huff factor is given in Ref.~\cite{Suzuki1987-aq} for most elements;
however, it is provided only as a function of the atomic number ($ Z $), neglecting isotope dependence.
The neglect of isotope dependence can introduce uncertainties in the deduced capture rates.
This issue is expected to become more pronounced in the future with the advent of the spectroscopy of radioactive muonic atoms~\cite{Strasser2005-hg,Strasser2009-ir,Antwis2025-rs}, which are muonic atoms formed by unstable nuclei.
\par
The Huff factor is defined as the ratio of the integral of the DIO electron energy spectrum to the decay rate of a muon in vacuum [see Eq.~\eqref{eq:huff_definition}].
Given that the electron energy spectrum is difficult to measure experimentally with high precision and that performing measurements for all nuclides is unrealistic, a theoretical calculation becomes indispensable.
The electron energy spectrum is theoretically obtained by first deriving the Coulomb potential from the charge distribution of the nucleus,
then obtaining the muon wave function by solving the Dirac equation,
and finally describing the muon decay using quantum electrodynamics (QED).
Among these processes, the dominant source of uncertainty in the calculated Huff factor stems from the treatment of the nuclear charge distribution.
Past studies often relied on simplified forms of the charge distribution, such as a two-parameter Fermi function,
without considering nuclear deformation.
In Refs.~\cite{Huff1961-ur,Suzuki1987-aq}, the details of the calculation regarding the charge distribution are not described, which prevents the consistent calculation of Huff factors for unlisted isotopes.
To address these issues, we provide a complete table of theoretically derived Huff factors based on the most reliable theoretical framework currently available.
\section{Theoretical framework}
\label{sec:method}
\par
We calculate the Huff factor in three steps.
First, the proton density distribution is obtained using a self-consistent Hartree-Fock plus Bardeen-Cooper-Schrieffer ($ \text{HF} + \text{BCS} $) calculation
including pairing and deformation, as described in Sect.~\ref{sec:p-density}.
Second, the nuclear charge distribution is deduced from the proton density distribution obtained above and the proton form factor, 
and the Coulomb potential is subsequently derived, as described in Sect.~\ref{sec:c-density}.
Finally, the energy spectrum of the emitted electrons in DIO is calculated by solving the Dirac equation with distortion effects included, as described in Sect.~\ref{sec:e-spectra}.
\par
We calculated the Huff factor for isotopes with atomic numbers ($ Z $) in the range $ 6 \leq Z \leq 94 $,
following the isotope set selected in the recent nuclear data evaluation of the nuclear muon capture rate~\cite{Iwamoto2025-cf}.
Light nuclei with $ Z < 6 $ are omitted in the present study because mean-field calculations, such as $ \text{HF} + \text{BCS} $ one, become less reliable in light nuclei,
and the convergence of the Huff factor with respect to the partial wave expansion is very poor for these nuclei.
Since the experimental uncertainties in $ \tau_{\urm{total}} $ are typically at the $ 1 \, \% $ level,
we set a target numerical precision of $ 0.1\, \% $ for the calculated Huff factor to ensure that the theoretical uncertainty remains negligible in the extraction of $ \Lambda_{\urm{cap}} $.
Note that, throughout this paper, we use the units of $ 4 \pi \epsilon_0 = c = \hbar = 1 $.
\subsection{Proton density distribution}
\label{sec:p-density}
\par
The nucleon density distributions are obtained through the fully self-consistent mean-field calculation.
We employ the Skyrme $ \text{HF} + \text{BCS} $ method represented
in the three-dimensional Cartesian coordinate space to calculate the ground states of nuclei,
considering nuclear pairing and deformation~\cite{
  Ebata.PhysRevC.111.014313}. 
In this work, the nuclear density distributions of even-even nuclei are calculated.
For nuclei with odd numbers of protons or neutrons,
the density distributions are approximated by averaging those of the neighboring even-even nuclei.
The mean-field methods for describing odd nuclei include the blocking method and the equal filling approximation (EFA). 
The EFA provides an excellent description of time-reversal-even components of the energy density functional~\cite{PhysRevC.78.014304}, such as the density in Eq.~\eqref{eq:dens}. 
The averaging procedure adopted here therefore provides a sufficiently good approximation for odd nuclei.
\par
The intrinsic proton density is written as
\begin{equation}
  \rho_p \left( \ve{r} \right)
  =
  \sum_{i, \sigma}
  n_i^p
  \left| \phi_i^p \left( \ve{r}, \sigma \right) \right|^2,
  \label{eq:dens}
\end{equation}
where $ \sigma $ denotes a spin,
and 
$ n_i^p \in \left[ 0, 1 \right] $ is a occupation probability of the proton single-particle state $ \phi_i^p $ labeled with $ i $.
The spatial degree of freedom is included only in the $ \phi_i^p $. 
They are obtained through the calculation of the HF and BCS gap equations iteratively.
The SkM* effective interaction~\cite{
  Bartel1982Nucl.Phys.A386_79}
is used in the calculation.
\par
To transform the intrinsic proton density distributions into the density in the laboratory frame, we average them over the solid angle as 
\begin{equation}
  \overline{\rho}_p
  \left( r \right)
  \equiv
  \frac{1}{4 \pi}
  \int
  \rho_p \left( \ve{r} \right)
  \, 
  d \Omega.
\end{equation}
\subsection{Charge density and Coulomb potential}
\label{sec:c-density}
\par
\begin{figure*}[tb]
  \centering
  \includegraphics[width=1.0\linewidth]{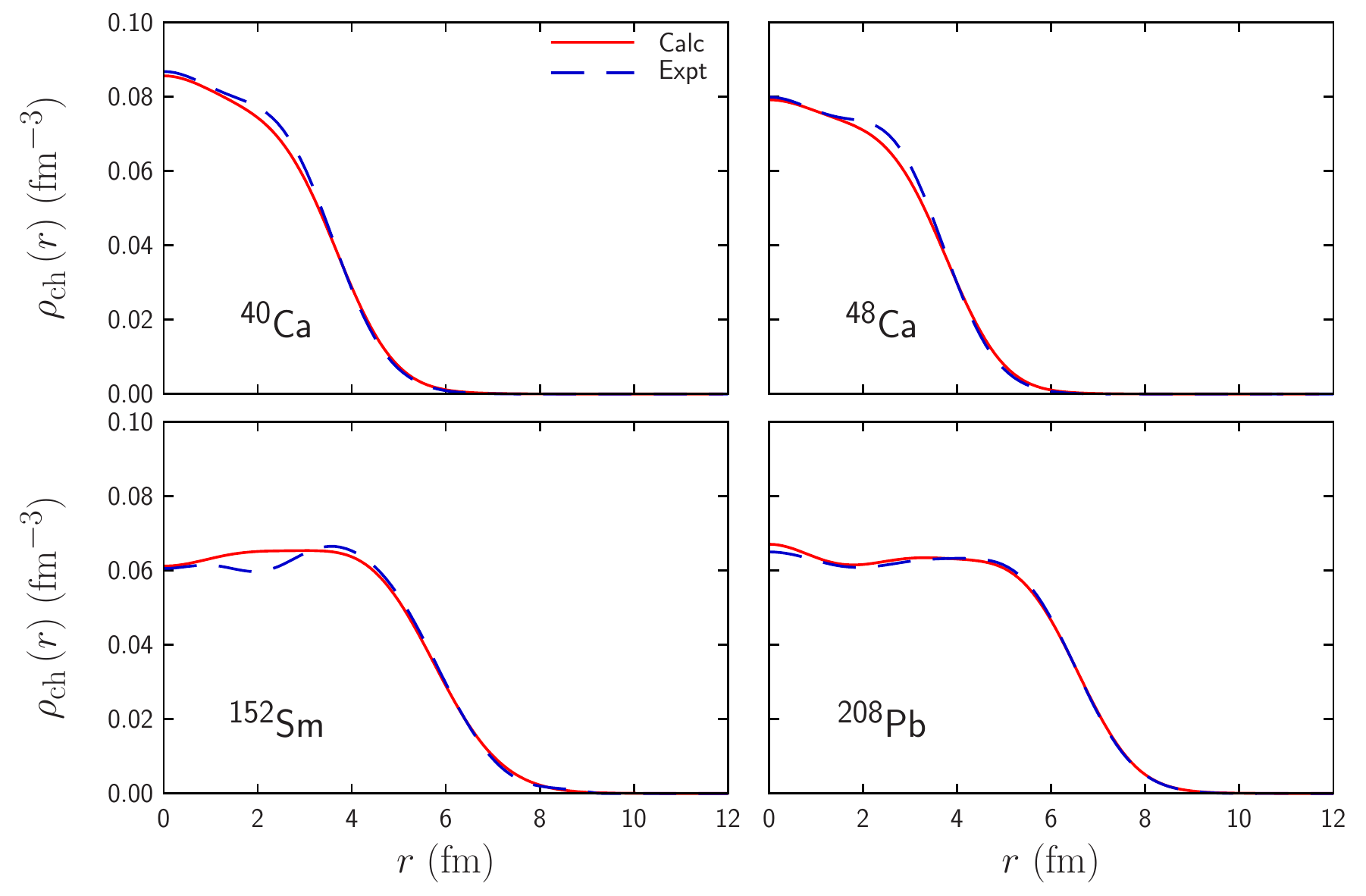}
  \caption{The calculated charge densities of $ \nuc{Ca}{40}{} $, $ \nuc{Ca}{48}{} $, $ \nuc{Sm}{152}{} $, and $ \nuc{Pb}{208}{} $
    compared with 
    the experimental data obtained by the electron scattering~\cite{
      DeVries1987At.DataNucl.DataTables36_495}.}
  \label{fig:chargeden}
\end{figure*}

Once the spherically averaged proton density distribution $ \overline{\rho}_p $ is obtained in the last section,
the charge density distribution $ \rho_{\urm{ch}} $ is calculated by using the convolution
\begin{equation}
  \label{eq:charge}
  \tilde{\rho}_{\urm{ch}} \left( q \right)
  =
  \tilde{\overline{\rho}}_p \left( q \right)
  \tilde{G}_{\urm{E} p} \left( q^2 \right),
\end{equation}
where $ \tilde{\rho}_{\urm{ch}} $ and $ \tilde{\overline{\rho}}_p $ are, respectively, the Fourier transform of $ \rho_{\urm{ch}} $ and $ \overline{\rho}_p $.
Here, $ \tilde{G}_{\urm{E} p} $ is the electric form factor of protons
and
we use $ \tilde{G}_{\urm{E} p} $ given by the Friedrich and Walcher~\cite{
  Friedrich2003Eur.Phys.J.A17_607}
\begin{align}
  \tilde{G}_{\urm{E} p} \left( q^2 \right)
  & =
    \frac{a_{10}}{\left( 1 + q^2 / a_{11} \right)^2}
    +
    \frac{a_{20}}{\left( 1 + q^2 / a_{21} \right)^2}
    \notag \\
  & \quad
    +
    a_{\urm{b}} q^2
    \left[
    \exp
    \left( - \frac{1}{2} \left( \frac{q - q_{\urm{b}}}{\sigma_{\urm{b}}} \right)^2 \right)
    +
    \exp
    \left( - \frac{1}{2} \left( \frac{q + q_{\urm{b}}}{\sigma_{\urm{b}}} \right)^2 \right)
    \right]
\end{align}
with
$ a_{10} =  1.041 $, 
$ a_{11} =  0.765 \, \mathrm{GeV}^2 / c^2 $,
$ a_{20} = -0.041 $,
$ a_{21} =  6.2   \, \mathrm{GeV}^2 / c^2 $,
$ a_{\urm{b}}  = -0.23 \, c^2 / \mathrm{GeV}^2 $,
$ q_{\urm{b}}  =  0.07 \, \mathrm{GeV} / c $,
and 
$ \sigma_{\urm{b}}  =  0.27 \, \mathrm{GeV} / c $.
We neglect the tiny contribution of the neutron electric form factor and nucleon magnetic form factors~\cite{
  Naito2021Phys.Rev.C104_024316,
  Naito2023Phys.Rev.C107_054307}.
The calculated charge densities of $ \nuc{Ca}{40}{} $, $ \nuc{Ca}{48}{} $, $ \nuc{Sm}{152}{} $, and $ \nuc{Pb}{208}{} $
are shown in Fig.~\ref{fig:chargeden}
as examples.
It can be seen that calculated charge densities reproduce
the experimental data obtained by the electron scattering~\cite{
  DeVries1987At.DataNucl.DataTables36_495}
well from light to heavy region for both spherical and deformed nuclei.
\par
The Coulomb potential formed by an atomic nucleus 
is calculated by~\cite{
  Minato2023Phys.Rev.C107_054314}
\begin{equation}
  \label{eq:Coul}
  V \left( r \right)
  =
  -
  \frac{4 \pi e^2}{r}
  \int_0^r
  \rho_{\urm{ch}} \left( r' \right)
  r'^2
  \, dr'
  -
  4 \pi e^2
  \int_r^{\infty}
  \rho_{\urm{ch}} \left( r' \right)
  r'
  \, dr'.
\end{equation}
Note that
the normal Coulomb potential $ -Z e^2/r $ is obtained
if one substitutes
$ \rho_{\urm{ch}} \left( \ve{r} \right) = Z \delta \left( \ve{r} \right) $.
\subsection{Electron spectra and Huff factor}
\label{sec:e-spectra}
\par
Using the partial wave expansion, the wave function of the emitted electron is written as
\begin{equation}
  \psi_{e, \ve{p}}^{s_e} \left( \ve{r} \right)
  =
  4 \pi
  \sum_{\kappa, \nu, m}
  i^{l_\kappa}
  \left( l_{\kappa}, m, 1/2, s_e \mid j_{\kappa}, \nu \right) 
  Y_{l_{\kappa}}^{m*} \left( \hat{\ve{p}} \right)
  e^{- i \delta_{\kappa}}
  \psi_{E}^{\kappa, \nu} \left( \ve{r} \right),
\end{equation}
where $\delta_\kappa$ is the phase shift to satisfy the incoming wave boundary condition.
Here, $ \kappa $ is a nonzero integer,
$ \left( l_{\kappa}, m, 1/2, s_e \mid j_{\kappa}, \nu \right) $ is a Clebsch-Gordan coefficient,
and the corresponding total angular momentum $ j_{\kappa} $ and orbital angular momentum $ l_{\kappa} $ are, respectively, given by
\begin{align}
  j_{\kappa}
  & = 
    \left| \kappa \right| - \frac{1}{2}, \\
  l_{\kappa}
  & = 
    j_{\kappa} + \frac{1}{2} \frac{\kappa}{\left| \kappa \right|}.
\end{align}
For a spherically symmetric Coulomb potential, a wave function with index $ \kappa $ can be written as
\begin{equation}
  \psi_E^{\kappa, \nu} \left( \ve{r} \right)
  =
  \begin{pmatrix}
    g_E^{\kappa}  \left( r \right) \chi_{\kappa}^{\nu}  \left( \hat{\ve{r}} \right) \\
    if_E^{\kappa} \left( r \right) \chi_{-\kappa}^{\nu} \left( \hat{\ve{r}} \right)
  \end{pmatrix},
\end{equation}
where $ g_E^{\kappa} \left( r \right) $ and $ f_E^{\kappa} \left( r \right) $ satisfy the radial Dirac equation
\begin{equation}
  \label{eq:Dirac_eq}
  \begin{pmatrix}
    \frac{d}{dr} + \frac{1 + \kappa}{r} & - E - m_{\ell} + V \left( r \right) \\
    E - m_{\ell} - V \left( r \right) & \frac{d}{dr} + \frac{1 - \kappa}{r}
  \end{pmatrix}
  \begin{pmatrix}
    g^{\kappa}_E \left( r \right) \\
    f^{\kappa}_E \left( r \right)
  \end{pmatrix}
  =
  0,
\end{equation}
with $m_\ell=m_e$ for the electron.
To solve the radial Dirac equations for the scattering states, we employ a fourth-order Runge-Kutta method.
For each $\kappa$, the regular solution at the origin is constructed from the small-$r$ behavior of the Dirac equation in the finite-size Coulomb potential and integrated outward up to a sufficiently large $r$, where the potential is well approximated by the point-Coulomb form, $-Ze^2/r$.
The numerical solution is then matched to the asymptotic Coulomb scattering solution at large $r$, and this matching determines the phase shift $ \delta_{\kappa} $ and the overall normalization.
Here, we employ the following normalization condition for the scattering states:
\begin{equation}
  \int
  \psi_{E'}^{\kappa', \nu' \dagger} \left( \ve{r} \right)
  \psi_E^{\kappa,\nu} \left( \ve{r} \right)
  \, d \ve{r} 
  =
  2 \pi
  \delta \left( E - E' \right)
  \delta_{\kappa, \kappa'}
  \delta_{\nu, \nu'}.
  \label{eq:normalization_scattering}
\end{equation}
\par
The muon wave function in the $ 1s $ orbital is given by
\begin{equation}
  \psi_{\mu}^{1s} \left( \ve{r} \right)
  =
  \begin{pmatrix}
    g_{\mu}  \left( r \right) \chi_{-1}^{s_\mu} \left( \hat{\ve{r}} \right) \\
    if_{\mu} \left( r \right) \chi_{+1}^{s_\mu} \left( \hat{\ve{r}} \right)
  \end{pmatrix}.
\end{equation}
The radial wave functions satisfy Eq.~\eqref{eq:Dirac_eq} with $ \kappa = -1 $ and $ m_{\ell} = m_{\mu} $.
For the bound state in the finite-size nuclear potential, the radial functions are required to be regular at the origin and to vanish at large $r$.
The normalization is given by
\begin{equation}
  \int_0^{\infty}
  \left[
    \left\{ g_{\mu} \left( r \right) \right\}^2
    +
    \left\{ f_{\mu} \left( r \right) \right\}^2
  \right]
  r^2
  \, dr
  =
  1.
\end{equation}
For the bound muon state, the same Runge-Kutta method is used, and the binding energy is determined by searching for the energy eigenvalue $E$ for which the radial wave function is normalizable.
In numerical calculations, the radial step size and the integration range were refined until the resultant Huff factor converged sufficiently.
\par
The formulation of the DIO spectrum is provided in Refs.~\cite{Haenggi:1974hp,Watanabe:1987su,Watanabe:1993emp,Czarnecki:2011mx}.
The spectrum is expressed as the sum over contributions labeled by the index $ \kappa $ as 
\begin{equation}
  \label{eq:DIO_spectrum_kappa_sum}
  \frac{d \Gamma}{d E_e}
  =
  \sum_{\kappa}
  \frac{d \Gamma_{\kappa}}{d E_e}.
\end{equation}
Each contribution of $ \kappa $ is defined by
\begin{align}
  \frac{d \Gamma_{\kappa}}{d E_e}
  & = 
    \frac{G_F^2}{48 \pi^4}
    \left( 2 j_{\kappa} + 1 \right)
    \sum_{J = \left| \kappa \right| - 1}^{\left| \kappa \right|}
    \int_0^{K_0}
    dK
    \,
    K^2
    \notag \\
  & \quad
    \times
    \left[
    K^2
    \left\{
    \left| S_{J, \kappa} \left(K, E_e \right) \right|^2
    +
    \frac{\left| S_{J, \kappa}^{+1} \left( K, E_e \right) + S_{J, \kappa}^{-1} \left( K, E_e \right) \right|^2}{\left( 2J+1 \right)^2}
    \right\}
    \right.
    \notag \\
  & \quad
    +
    \left.
    \left( K_0^2 - K^2 \right)
    \right.
    \notag \\
  & \quad
    \times
    \left.
    \left\{
    \frac{\left| S_{J, \kappa}^0 \left( K, E_e \right) \right|^2}{J \left( J + 1 \right)}
    +
    \frac{\left| S_{J, \kappa}^{+1} \left( K, E_e \right) \right|^2}{\left( J + 1 \right) \left( 2J + 1 \right)}
    +
    \frac{\left| S_{J, \kappa}^{-1} \left( K, E_e \right) \right|^2}{J \left( 2J + 1 \right)}
    \right\}
    \right.
    \notag \\
  & \quad
    \left.
    +
    \frac{2 K_0 K}{2J + 1}
    \Re
    \left\{ S_{J, \kappa}^* \left( K, E_e \right)
    \left[
    S_{J, \kappa}^{+1} \left( K, E_e \right)
    +
    S_{J, \kappa}^{-1} \left( K, E_e \right)
    \right]
    \right\}
    \right]
    \label{eq:DIO_spectrum}
\end{align}
for $ m_e < E_e < m_{\mu} - B_{\mu} $
with $ K_0 = m_{\mu} - B_\mu - E_e $
and
\begin{align}
  & S_{J,\kappa} \left( K, E_e \right)
    \notag \\
  & =
    \frac{1}{2}
    \int_0^{\infty}
    dr
    \, 
    r^2 j_J \left( Kr \right) \notag \\
  & \quad
    \times
    \left[
    \left\{
    1 + \left( -1 \right)^{l_{\kappa} + J}
    \right\}
    \left\{
    g_{E_e}^{\kappa} \left( r \right)
    g_{\mu} \left( r \right)
    +
    f_{E_e}^{\kappa} \left( r \right)
    f_{\mu} \left( r \right)
    \right\}
    \right.
    \notag \\
  & \qquad
    \left.
    +
    i
    \left\{
    1 - \left( -1 \right)^{l_{\kappa} + J}
    \right\}
    \left\{
    f_{E_e}^{\kappa} \left( r \right)
    g_{\mu} \left( r \right)
    -
    g_{E_e}^{\kappa} \left( r \right)
    f_{\mu} \left( r \right)
    \right\}
    \right], \\
  & S_{J,\kappa}^0 \left( K, E_e \right)
    \notag \\
  & =
    \frac{1}{2}
    \int_0^{\infty}
    dr
    \, 
    r^2 j_J \left( Kr \right)
    \notag \\
  & \quad
    \times
    \left[
    \left\{
    1 + \left( -1 \right)^{l_{\kappa} + J}
    \right\}
    \left( 1 + \kappa \right)
    \left\{
    g_{E_e}^{\kappa} \left( r \right)
    g_{\mu} \left( r \right)
    -
    f_{E_e}^{\kappa} \left( r \right)
    f_{\mu} \left( r \right)
    \right\}
    \right.
    \notag \\
  & \qquad
    \left.
    +
    i
    \left\{
    1 - \left( -1 \right)^{l_{\kappa} + J}
    \right\}
    \left( 1 - \kappa \right)
    \left\{
    f_{E_e}^{\kappa} \left( r \right)
    g_{\mu} \left( r \right)
    +
    g_{E_e}^{\kappa} \left( r \right)
    f_{\mu} \left( r \right)
    \right\}
    \right], \\
  & S_{J, \kappa}^{+1} \left( K, E_e \right)
    \notag \\
  & =
    \frac{1}{2}
    \int_0^{\infty}
    dr
    \, 
    r^2 j_{J+1} \left( Kr \right)
    \notag \\
  & \quad
    \times
    \left[
    \left\{
    1 + \left( -1 \right)^{l_{\kappa} + J}
    \right\}
    \right.
    \notag \\
  & \qquad
    \times \left\{
    \left( 2 + J + \kappa \right)
    g_{E_e}^{\kappa} \left( r \right)
    f_{\mu} \left( r \right)
    -
    \left( J - \kappa \right)
    f_{E_e}^{\kappa} \left( r \right)
    g_{\mu} \left( r \right)
    \right\}
    \notag \\
  & \quad
    +
    i
    \left\{
    1 - \left( -1 \right)^{l_{\kappa} + J}
    \right\}
    \notag \\
  & \qquad
    \times
    \left.
    \left\{
    \left( 2 + J - \kappa \right)
    f_{E_e}^{\kappa} \left( r \right)
    f_{\mu} \left( r \right)
    +
    \left( J + \kappa \right)
    g_{E_e}^{\kappa} \left( r \right)
    g_{\mu} \left( r \right)
    \right\}
    \right],
\end{align}
\begin{align}
  & S_{J, \kappa}^{-1} \left( K, E_e \right)
    \notag \\
  & =
    \frac{1}{2}
    \int_0^{\infty}
    dr
    \,
    r^2 j_{J - 1} \left( Kr \right)
    \notag \\
  & \quad
    \times
    \left[
    \left\{
    1 + \left( -1 \right)^{l_{\kappa} + J}
    \right\}
    \right.
    \notag \\
  & \qquad
    \times
    \left\{
    \left( 1 - J + \kappa \right)
    g_{E_e}^{\kappa} \left( r \right)
    f_{\mu} \left( r \right)
    +
    \left( 1 + J + \kappa \right)
    f_{E_e}^{\kappa} \left( r \right)
    g_{\mu} \left( r \right)
    \right\}
    \notag \\
  & \quad
    +
    i
    \left\{
    1 - \left( -1 \right)^{l_{\kappa} + J}
    \right\}
    \notag\\
  & \qquad
    \left.
    \times
    \left\{
    \left( 1 - J - \kappa \right)
    f_{E_e}^{\kappa} \left( r \right)
    f_{\mu} \left( r \right)
    -
    \left( 1 + J - \kappa \right)
    g_{E_e}^{\kappa} \left( r \right)
    g_{\mu} \left( r \right)
    \right\}
    \right].
\end{align}
Here, $ G_F = 1.166\times 10^{-5} \, \mathrm{GeV}^{-2} $ is the Fermi coupling constant, and $ j_J $ is the spherical Bessel function of the $ J $-th order.
In Eq.~\eqref{eq:DIO_spectrum},
the terms including $ S_{J,\kappa}^0 \left( K, E_e \right) $ or $ S_{J,\kappa}^{-1} \left( K, E_e \right) $ are set to zero for $ J = 0 $.
\par
The Huff factor is given by
\begin{equation}
  \label{eq:huff_definition}
  Q
  =
  \frac{1}{\Gamma_0}
  \int_{m_e}^{m_{\mu} - B_{\mu}}
  \frac{d \Gamma}{d E_e}
  \,
  dE_e,
\end{equation}
where $\Gamma_0$ is the decay width of the free muon, given by
\begin{equation}
  \Gamma_0
  =
  \frac{G_F^2 m_{\mu}^5}{192 \pi^3}f\left(\frac{m_e}{m_\mu}\right).
\end{equation}
The function
\begin{equation}
  f \left( \delta \right) = 1 - 8\delta^2 - 24\delta^4\log \delta + 8\delta^6 - \delta^8
\end{equation}
accounts for the electron-mass correction~\cite{Pak:2008qt,Czarnecki:2025phw} and evaluates to
$ f \left( m_e / m_{\mu} \right) \simeq 0.99981 $.
This theoretical expression respects the requirement that $ Q $ tends to $ 1 $ as $ Z $ approaches $ 0 $.
\par
In the numerical calculation of the Huff factors, it is necessary to truncate the summation over $ \kappa $.
We perform the summation over the range $ -\kappa_{\urm{max}} \le \kappa \le \kappa_{\urm{max}} $, 
where the cutoff parameter $ \kappa_{\urm{max}} $ is chosen so that the truncation error is less than $ 0.002 \, \% $.
A larger value of $ \kappa_{\urm{max}} $ is required for light nuclei,
and the values of $ \kappa_{\urm{max}} $ employed in the calculation are listed together with the results in \ref{sec:app_result}.
Further details on the convergence are given in \ref{app:kappa_convergence}.
\par
Several higher-order effects on the DIO spectrum have been discussed in previous studies, including nuclear-recoil effects~\cite{Haenggi:1974hp,Czarnecki:2011mx,Shanker:1981mi,Shanker:1996rz,Heeck:2021adh,Kaygorodov:2025yag,Fontes:2025mps}, vacuum-polarization effects~\cite{Haenggi:1974hp,Szafron:2016cbv,Heeck:2021adh,Fontes:2024yvw,Kaygorodov:2025yag,Fontes:2025mps}, and photonic corrections, such as those associated with collinear or soft photon radiation~\cite{Czarnecki:2014cxa,Szafron:2015kja,Szafron:2016cbv,Heeck:2021adh,Fontes:2024yvw,Fontes:2025mps}.
Although these effects can modify the differential spectral shape, their impact on the integrated decay rate relevant to the Huff factor is expected to be much smaller.
For this reason, we do not include them in the present calculation.
\par
The nuclear-recoil and vacuum-polarization effects reduce the endpoint energy and suppress the spectrum near the high-energy endpoint, which is particularly important for precision estimates of the background to $\mu^- \to e^-$ conversion.
However, this region contributes only a tiny fraction to the total decay rate.
Photonic corrections are known to generate large logarithmic terms that affect the spectrum not only near the endpoint but also around the peak.
For inclusive decay rates, however, these logarithms cancel according to the Kinoshita--Lee--Nauenberg theorem~\cite{Czarnecki:2014cxa,Kinoshita:1962ur,Lee:1964is}.
Moreover, if the photonic corrections approximately factorize in the decay rate, their effects are expected to be further partially canceled in the ratio of the bound-muon and free-muon decay rates that defines the Huff factor.
We therefore do not incorporate photonic corrections in the present work.
\section{Result and discussion}
\begin{table*}[ht]
\centering
  \caption{Effective Huff factors for each element, calculated as the weighted average of the Huff factors using natural abundance~\cite{natural_abundance}. For convenience, the same dataset is provided as Supplementary Material in CSV format.}
  \label{tab:average_huff_factors}
  \begin{tabular}{ccD{.}{.}{4}@{\hspace{4.5em}}ccD{.}{.}{4}@{\hspace{4.5em}}ccD{.}{.}{4}}
\toprule
    $ Z $ & Element & \multicolumn{1}{l}{Huff factor} & $ Z $ & Element & \multicolumn{1}{l}{Huff factor} & $ Z $ & Element & \multicolumn{1}{l}{Huff factor} \\
\midrule
6 & C & 0.9989 & 33 & As & 0.9558 & 62 & Sm & 0.8879 \\
7 & N & 0.9984 & 34 & Se & 0.9535 & 63 & Eu & 0.8858 \\
8 & O & 0.9979 & 35 & Br & 0.9509 & 64 & Gd & 0.8848 \\
9 & F & 0.9972 & 36 & Kr & 0.9486 & 65 & Tb & 0.8826 \\
10 & Ne & 0.9965 & 37 & Rb & 0.9460 & 66 & Dy & 0.8808 \\
11 & Na & 0.9956 & 38 & Sr & 0.9436 & 67 & Ho & 0.8787 \\
12 & Mg & 0.9947 & 39 & Y & 0.9409 & 68 & Er & 0.8766 \\
13 & Al & 0.9936 & 40 & Zr & 0.9384 & 69 & Tm & 0.8745 \\
14 & Si & 0.9924 & 41 & Nb & 0.9359 & 70 & Yb & 0.8728 \\
15 & P & 0.9912 & 42 & Mo & 0.9334 & 71 & Lu & 0.8706 \\
16 & S & 0.9898 & 44 & Ru & 0.9285 & 72 & Hf & 0.8687 \\
17 & Cl & 0.9884 & 45 & Rh & 0.9259 & 73 & Ta & 0.8665 \\
18 & Ar & 0.9869 & 46 & Pd & 0.9238 & 74 & W & 0.8643 \\
19 & K & 0.9852 & 47 & Ag & 0.9212 & 75 & Re & 0.8621 \\
20 & Ca & 0.9835 & 48 & Cd & 0.9190 & 76 & Os & 0.8601 \\
21 & Sc & 0.9817 & 49 & In & 0.9165 & 77 & Ir & 0.8577 \\
22 & Ti & 0.9798 & 50 & Sn & 0.9141 & 78 & Pt & 0.8555 \\
23 & V & 0.9779 & 51 & Sb & 0.9119 & 79 & Au & 0.8533 \\
24 & Cr & 0.9758 & 52 & Te & 0.9099 & 80 & Hg & 0.8512 \\
25 & Mn & 0.9737 & 53 & I & 0.9073 & 81 & Tl & 0.8493 \\
26 & Fe & 0.9715 & 54 & Xe & 0.9043 & 82 & Pb & 0.8473 \\
27 & Co & 0.9694 & 55 & Cs & 0.9027 & 83 & Bi & 0.8454 \\
28 & Ni & 0.9669 & 56 & Ba & 0.9006 & 90 & Th & 0.8370 \\
29 & Cu & 0.9649 & 57 & La & 0.8982 & 91 & Pa & 0.8347 \\
30 & Zn & 0.9626 & 58 & Ce & 0.8957 & 92 & U & 0.8344 \\
31 & Ga & 0.9604 & 59 & Pr & 0.8932 &  & &  \\
32 & Ge & 0.9581 & 60 & Nd & 0.8912 &  & &  \\
\bottomrule
\end{tabular}
\end{table*}

\par
The complete set of calculated Huff factors for all isotopes considered in this work is summarized in~\ref{sec:app_result}.
These values provide a consistent dataset obtained within a single theoretical framework given in Sect.~\ref{sec:method},
offering a basis for both a systematic understanding of the nuclear muon capture rate and reliable use in applications.
\par
To assess the nuclear-structure model dependence associated with the choice of Skyrme parameterization, we performed additional calculations of the Huff factor for $ {}^{40}\mathrm{Ca} $ and $ {}^{208}\mathrm{Pb} $ using the SLy4 effective interaction~\cite{
  Chabanat1998Nucl.Phys.A635_231},
in addition to the SkM* one used in the present work.
The calculated values are $ 0.9835 $ (SkM*) and $ 0.9835 $ (SLy4) for $ {}^{40}\mathrm{Ca} $, and $ 0.8474 $ (SkM*) and $ 0.8476 $ (SLy4) for $ {}^{208}\mathrm{Pb} $.
The differences are well within our target numerical precision of $ 0.1 \, \% $, indicating that the dependence on the choice of Skyrme parameterization is negligible.
\par
To further examine the uncertainty arising from the nuclear charge distribution, we performed additional calculations using charge densities derived from electron scattering data in the Fourier-Bessel analysis for $ {}^{40}\mathrm{Ca} $, $ {}^{48}\mathrm{Ca} $, $ {}^{152}\mathrm{Sm} $, and $ {}^{208}\mathrm{Pb} $, as shown in Fig.~\ref{fig:chargeden}.
The resultant Huff factors of $ 0.9834 $, $ 0.9834 $, $ 0.8886 $, and $ 0.8472 $ for these four nuclei, respectively, agree with those from the calculated charge distributions to within $ 0.1\, \% $, indicating that our calculation scheme to obtain charge distributions is reliable at the present level of accuracy.
This supports the use of a unified theoretical framework applicable to all nuclei, for which experimental charge densities are not universally available.
\par
The Huff factor exhibits a monotonic decrease as the atomic number $ Z $ increases.
This trend is consistent with the stronger Coulomb field in heavier nuclei, which suppresses the muon decay rate.
The treatment of the lepton wave functions in the strong Coulomb field is crucial for an accurate evaluation of the Huff factor.
To clarify the role of Coulomb distortion on the emitted electron, in~\ref{app:distortion} we derive an analytic expression based on the plane wave approximation and compare it with the results obtained in this work.
As shown in Appendix~\ref{app:distortion}, the plane wave approximation does not reproduce the quantitative $Z$-dependence of the Huff factor, whereas the inclusion of the Coulomb distortion of the emitted electron restores the much milder decrease obtained in the full calculation.
\par
In addition to this global behavior, the isotope dependence of the Huff factor is revealed for the first time in this study.
While our results show slight increases of the Huff factor with increasing mass number $ A $, the magnitude of this isotope dependence is found to be small.
No isotope dependence is observed for light nuclei within the $ 0.1 \, \% $ precision targeted in the present work,
and the deviations remain within approximately $ 0.1 \, \% $ even for heavy nuclei.
Although previous calculations neglected this isotope dependence, our results indicate that the assumption of a constant Huff factor for a given element is sufficiently accurate for stable nuclei.
\par
Table~\ref{tab:average_huff_factors} provides the effective Huff factor for each element, evaluated for natural abundant isotopes.
These values are calculated as the weighted average of the Huff factors for each isotope using natural abundances~\cite{natural_abundance}.
Owing to the minimal isotope dependence, this effective value offers practical utility for the derivation of $ \Lambda_{\urm{cap}} $ from the measured $ \tau_{\urm{total}} $ in natural-target experiments.
\begin{figure}[tb]
  \centering
  \includegraphics[width=1.0\linewidth]{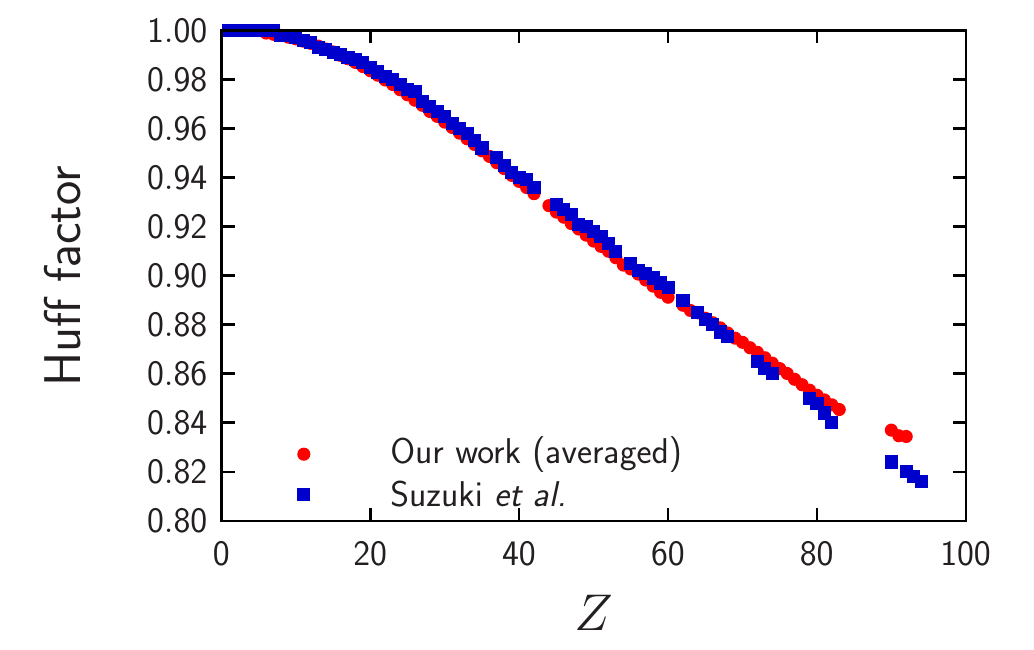}
  \caption{The calculated Huff factors for each element obtained in the present study given
    in Table~\ref{tab:average_huff_factors} (red circles) and those in Ref.~\cite{Suzuki1987-aq} (blue squares).}
  \label{fig:huff_suzuki}
\end{figure}
\begin{table}[t]
  \caption{
    Comparisons of the obtained Huff factors for specific nuclei with previous studies~\cite{Suzuki1987-aq,Watanabe:1993emp,Czarnecki:2011mx,Zinatulina2019-nv,Blair1962-kp}.
    The values in the Present column are a mixture of the calculated Huff factors for the respective isotopes (indicated with mass number, obtained from \ref{sec:app_result}) and the averaged one for each element (indicated by element symbol only, obtained from Table~\ref{tab:average_huff_factors}).
    All values from Ref.~\cite{Suzuki1987-aq} are the Huff factor for each element, as isotope dependence was not considered in their work.
    All values from Ref.~\cite{Czarnecki:2025phw} are renormalized to the definition of the Huff factor adopted in the present work; the original values are divided by 0.99981 (see text for details).}
  \label{tab:comparison}
  \centering
  \begin{tabular}{rD{.}{.}{4}l}
    \toprule
    Nucleus & \multicolumn{2}{c}{Huff factor} \\
            & \multicolumn{1}{c}{Present} & \multicolumn{1}{c}{Previous} \\
    \midrule
    $ \mathrm{C} $       & 0.9989 & $ 1.00 $~\cite{Suzuki1987-aq},  $ 0.9989 $~\cite{Czarnecki:2025phw} \\
    $ \mathrm{N} $       & 0.9984 & $ 1.00 $~\cite{Suzuki1987-aq},  $ 0.9984 $~\cite{Czarnecki:2025phw} \\
    $ \nuc{O}{16}{} $    & 0.9979 & $ 0.998 $~\cite{Suzuki1987-aq}, $ 0.994 $~\cite{Watanabe:1993emp}, $ 0.9978 $~\cite{Czarnecki:2025phw} \\
    $ \mathrm{F} $       & 0.9972 & $ 0.998 $~\cite{Suzuki1987-aq}, $ 0.9971 $~\cite{Czarnecki:2025phw} \\
    $ \nuc{Al}{27}{} $   & 0.9936 & $ 0.993 $~\cite{Suzuki1987-aq}, $ 0.992 $~\cite{Watanabe:1993emp}, $ 0.9934 $~\cite{Czarnecki:2011mx} \\
    $ \nuc{Si}{28}{} $   & 0.9924 & $ 0.992 $~\cite{Suzuki1987-aq}, $ 0.991 $~\cite{Watanabe:1993emp} \\
    $ \nuc{Ca}{40}{} $   & 0.9835 & $ 0.985 $~\cite{Suzuki1987-aq}, $ 0.981 $~\cite{Watanabe:1993emp} \\
    $ \mathrm{Ti} $      & 0.9798 & $ 0.981 $~\cite{Suzuki1987-aq}, $ 0.98 $~\cite{Zinatulina2019-nv} \\
    $ \mathrm{V} $       & 0.9779 & $ 0.980 $~\cite{Suzuki1987-aq}, $ 0.98 $~\cite{Blair1962-kp}\\
    $ \nuc{Fe}{56}{} $   & 0.9715 & $ 0.975 $~\cite{Suzuki1987-aq}, $ 0.98 $~\cite{Blair1962-kp}, $ 0.971 $~\cite{Watanabe:1993emp} \\
    $ \mathrm{Co} $      & 0.9694 & $ 0.971 $~\cite{Suzuki1987-aq}, $ 0.97 $~\cite{Blair1962-kp}\\
    $ \mathrm{Ni} $      & 0.9669 & $ 0.969 $~\cite{Suzuki1987-aq}, $ 0.97 $~\cite{Blair1962-kp}\\
    $ \mathrm{Zn} $      & 0.9626 & $ 0.965 $~\cite{Suzuki1987-aq}, $ 0.96 $~\cite{Blair1962-kp}\\
    $ \mathrm{Se} $      & 0.9535 & $ 0.955 $~\cite{Suzuki1987-aq}, $ 0.96 $~\cite{Zinatulina2019-nv} \\
    $ \mathrm{Kr} $      & 0.9486 & $ 0.95 $~\cite{Zinatulina2019-nv} \\
    $ \nuc{Zr}{90}{} $   & 0.9383 & $ 0.940 $~\cite{Suzuki1987-aq}, $ 0.936 $~\cite{Watanabe:1993emp} \\
    $ \nuc{Mo}{96}{} $   & 0.9334 & $ 0.936 $~\cite{Suzuki1987-aq}, $ 0.932 $~\cite{Watanabe:1993emp} \\
    $ \mathrm{Cd} $      & 0.9190 & $ 0.921 $~\cite{Suzuki1987-aq}, $ 0.92 $~\cite{Zinatulina2019-nv} \\
    $ \nuc{Sn}{118}{} $  & 0.9140 & $ 0.918 $~\cite{Suzuki1987-aq}, $ 0.914 $~\cite{Watanabe:1993emp}, $ 0.92 $~\cite{Blair1962-kp} \\
    $ \mathrm{Sm} $      & 0.8879 & $ 0.890 $~\cite{Suzuki1987-aq}, $ 0.89 $~\cite{Zinatulina2019-nv} \\
    $ \mathrm{W} $       & 0.8643 & $ 0.860 $~\cite{Suzuki1987-aq}, $ 0.85 $~\cite{Blair1962-kp}\\
    $ \nuc{Pb}{208}{} $  & 0.8474 & $ 0.844 $~\cite{Suzuki1987-aq}, $ 0.847 $~\cite{Watanabe:1993emp}, $ 0.84 $~\cite{Blair1962-kp} \\
    $ \nuc{Bi}{209}{} $  & 0.8454 & $ 0.840 $~\cite{Suzuki1987-aq}, $ 0.845 $~\cite{Watanabe:1993emp} \\
    \bottomrule
  \end{tabular}
\end{table}
\par
Figure~\ref{fig:huff_suzuki} shows the effective Huff factors listed in Table~\ref{tab:average_huff_factors} together with those reported in Suzuki~\textit{et al.}~\cite{Suzuki1987-aq}.
Our calculated values are slightly smaller for light nuclei and slightly larger for heavy nuclei.
Table~\ref{tab:comparison} summarizes the Huff factors for a representative set of elements and isotopes.
These values allow a direct comparison between the present work and earlier calculations among the representative nuclei.
\par
The electron spectra used in our calculation follow the formalism of Watanabe~\textit{et al.}~\cite{Watanabe:1993emp} (see Sect.~\ref{sec:e-spectra}), and only the difference between these two calculations lies in the treatment of the nuclear charge density; 
the present study obtains it from a microscopic calculation that incorporates nuclear deformation and pairing,
whereas Ref.~\cite{Watanabe:1993emp} used a two-parameter Fermi function as the nuclear charge density.
This improved description of the charge density accounts for the small deviations seen in Table~\ref{tab:comparison}.
The recent calculation by Czarnecki~\textit{et al.}~\cite{Czarnecki:2011mx} reports a Huff factor only for $ \nuc{Al}{27}{} $,
and our value of $ 0.9935 $ shows excellent agreement with their reported value of $ 0.9934 $.
\par
Complementary results for $ 4 \le Z \le 9 $ have recently been reported in Ref.~\cite{Czarnecki:2025phw}, where a point nucleus is assumed.
Because Ref.~\cite{Czarnecki:2025phw} adopts a different normalization from ours, their values are divided by
$ f \left( m_e / m_{\mu} \right) \simeq 0.99981 $
in Table~\ref{tab:comparison} to allow a proper comparison.
Our results for the lowest-$ Z $ nuclei considered here are consistent with theirs within the accuracy required for our purposes.
This agreement suggests that the effect of the finite nuclear charge distribution is small for light nuclei.
\par
Other reported values~\cite{Suzuki1987-aq,Zinatulina2019-nv,Blair1962-kp} did not provide their calculation procedures
and cited Huff's original work~\cite{Huff1961-ur} only,
leaving the source of the discrepancies unexplained.
In contrast, the present work derives all the isotopic and elemental Huff factors using a unified and fully specified framework, suggesting that the present values constitute the most robust and consistent dataset currently available.
\section{Summary and outlook}
\par
In this study, we have provided a comprehensive set of Huff factors for nuclei with $ 6 \leq Z \leq 94 $.
The calculations were performed within the most reliable theoretical framework currently available, providing the most consistent set of values to date.
\par
Our calculations confirmed the monotonic decrease of the Huff factor with increasing atomic number ($ Z $).
The isotope dependence, which had been neglected in most previous studies, was explicitly evaluated for the first time and found to be tiny.
Although the magnitude of this dependence is small, its systematic evaluation allowed us to derive effective Huff factors for each element, offering practical utility for analyzing experimental data obtained using natural targets.
Furthermore, the consistency of our results with previous benchmark calculations supports the robustness of the present framework.
\par
The complete set of Huff factors obtained in this work provides a necessary foundation for revising related nuclear data.
In particular, these values will be incorporated into an updated evaluation of the total muon capture rate ($ \Lambda_{\urm{cap}} $), which is essential for correcting and improving the existing nuclear data evaluation~\cite{Iwamoto2025-cf}.
This refinement represents an important step toward establishing a unified and reliable muon nuclear data library~\cite{Niikura2024-ty}.
% 
%% ============== Others ==================
% 
\section*{Declaration of Competing Interest}
\par
The authors declare that they have no known competing financial interests or personal relationships that could have appeared to influence the work reported in this paper.
\section*{Acknowledgment}
\par
We gratefully acknowledge helpful discussions with
Teiichiro~Matsuzaki (RIKEN),
Rurie~Mizuno (TRIUMF),
and Hiroki~Iwamoto (JAEA).
Y.~U.~acknowledges the JSPS KAKENHI Grant Numbers JP23K13106 and JP24K07061.
T.~N.~acknowledges
the JSPS Grant-in-Aid for Transformative Research Areas (A) under Grant No.~JP25H01558,
the JSPS Grant-in-Aid for Early-Career Scientists under Grant No.~JP24K17057,
the JSPS Grant-in-Aid for JSPS Fellows under Grant No.~JP25KJ0405,
the JSPS Grant-in-Aid for Scientific Research (S) under Grant No.~JP25H00402,
the JSPS Grant-in-Aid for Scientific Research (B) under Grant Nos.~JP23K26538 and JP25K01003,
the JSPS Grant-in-Aid for Scientific Research (C) under Grant No.~JP23K03426,
and
JST COI-NEXT Grant No.~JPMJPF2221.
The numerical calculations were partly performed on cluster computers at the RIKEN iTHEMS Center.

%% =============== Appendix ===================
% 
\appendix
\section{Comprehensive list of calculated Huff factors}
\label{sec:app_result}
\begin{table*}[p]
\centering
\caption{Calculated Huff factors and $ \kappa_{\urm{max}} $ for each isotope. For convenience, the same dataset is provided as Supplementary Material in CSV format.}
\label{tab:huff_factors_1}
\begin{tabular}{cccc@{\hspace{5em}}cccc@{\hspace{5em}}cccc}
\toprule
Z & A & Huff factor & $\kappa_\mathrm{max}$ & Z & A & Huff factor & $\kappa_\mathrm{max}$ & Z & A & Huff factor & $\kappa_\mathrm{max}$ \\
\midrule
6 & 12 & 0.9989 & 71 & 20 & 46 & 0.9835 & 22 & 30 & 67 & 0.9627 & 15 \\
6 & 13 & 0.9989 & 71 & 20 & 47 & 0.9835 & 22 & 30 & 68 & 0.9627 & 15 \\
6 & 14 & 0.9989 & 71 & 20 & 48 & 0.9835 & 22 & 30 & 69 & 0.9628 & 15 \\
7 & 13 & 0.9984 & 61 & 21 & 44 & 0.9817 & 21 & 30 & 70 & 0.9629 & 15 \\
7 & 14 & 0.9984 & 61 & 21 & 45 & 0.9817 & 21 & 30 & 71 & 0.9629 & 15 \\
7 & 15 & 0.9984 & 61 & 21 & 46 & 0.9817 & 21 & 30 & 72 & 0.9630 & 15 \\
8 & 16 & 0.9979 & 53 & 21 & 47 & 0.9817 & 21 & 31 & 67 & 0.9603 & 15 \\
8 & 17 & 0.9979 & 53 & 21 & 48 & 0.9817 & 21 & 31 & 68 & 0.9603 & 15 \\
8 & 18 & 0.9979 & 53 & 22 & 44 & 0.9798 & 20 & 31 & 69 & 0.9604 & 15 \\
9 & 18 & 0.9972 & 47 & 22 & 45 & 0.9798 & 20 & 31 & 70 & 0.9604 & 15 \\
9 & 19 & 0.9972 & 47 & 22 & 46 & 0.9798 & 20 & 31 & 71 & 0.9605 & 15 \\
10 & 20 & 0.9965 & 43 & 22 & 47 & 0.9798 & 20 & 32 & 68 & 0.9578 & 15 \\
10 & 21 & 0.9965 & 43 & 22 & 48 & 0.9798 & 20 & 32 & 69 & 0.9579 & 15 \\
10 & 22 & 0.9965 & 43 & 22 & 49 & 0.9798 & 20 & 32 & 70 & 0.9580 & 15 \\
11 & 22 & 0.9956 & 39 & 22 & 50 & 0.9798 & 20 & 32 & 71 & 0.9580 & 15 \\
11 & 23 & 0.9956 & 39 & 23 & 48 & 0.9778 & 19 & 32 & 72 & 0.9581 & 15 \\
12 & 24 & 0.9947 & 36 & 23 & 49 & 0.9778 & 19 & 32 & 73 & 0.9582 & 15 \\
12 & 25 & 0.9947 & 36 & 23 & 50 & 0.9778 & 19 & 32 & 74 & 0.9582 & 15 \\
12 & 26 & 0.9947 & 36 & 23 & 51 & 0.9779 & 19 & 32 & 75 & 0.9583 & 15 \\
13 & 26 & 0.9936 & 33 & 24 & 50 & 0.9758 & 19 & 32 & 76 & 0.9584 & 15 \\
13 & 27 & 0.9936 & 33 & 24 & 51 & 0.9758 & 19 & 33 & 73 & 0.9557 & 14 \\
14 & 28 & 0.9924 & 31 & 24 & 52 & 0.9758 & 19 & 33 & 74 & 0.9557 & 14 \\
14 & 29 & 0.9924 & 31 & 24 & 53 & 0.9758 & 19 & 33 & 75 & 0.9558 & 14 \\
14 & 30 & 0.9924 & 31 & 24 & 54 & 0.9759 & 19 & 34 & 74 & 0.9532 & 14 \\
14 & 31 & 0.9924 & 31 & 25 & 52 & 0.9737 & 18 & 34 & 75 & 0.9533 & 14 \\
14 & 32 & 0.9924 & 31 & 25 & 53 & 0.9737 & 18 & 34 & 76 & 0.9533 & 14 \\
15 & 31 & 0.9912 & 29 & 25 & 54 & 0.9737 & 18 & 34 & 77 & 0.9534 & 14 \\
15 & 32 & 0.9912 & 29 & 25 & 55 & 0.9737 & 18 & 34 & 78 & 0.9535 & 14 \\
15 & 33 & 0.9912 & 29 & 26 & 54 & 0.9715 & 17 & 34 & 79 & 0.9535 & 14 \\
16 & 32 & 0.9898 & 27 & 26 & 55 & 0.9715 & 17 & 34 & 80 & 0.9536 & 14 \\
16 & 33 & 0.9898 & 27 & 26 & 56 & 0.9715 & 17 & 34 & 81 & 0.9536 & 14 \\
16 & 34 & 0.9898 & 27 & 26 & 57 & 0.9716 & 17 & 34 & 82 & 0.9536 & 14 \\
16 & 35 & 0.9898 & 27 & 26 & 58 & 0.9716 & 17 & 35 & 79 & 0.9509 & 14 \\
16 & 36 & 0.9898 & 27 & 26 & 59 & 0.9717 & 17 & 35 & 80 & 0.9510 & 14 \\
17 & 35 & 0.9884 & 26 & 26 & 60 & 0.9717 & 17 & 35 & 81 & 0.9510 & 14 \\
17 & 36 & 0.9884 & 26 & 27 & 56 & 0.9692 & 17 & 36 & 78 & 0.9483 & 13 \\
17 & 37 & 0.9884 & 26 & 27 & 57 & 0.9693 & 17 & 36 & 79 & 0.9483 & 13 \\
18 & 36 & 0.9868 & 24 & 27 & 58 & 0.9693 & 17 & 36 & 80 & 0.9484 & 13 \\
18 & 37 & 0.9868 & 24 & 27 & 59 & 0.9694 & 17 & 36 & 81 & 0.9484 & 13 \\
18 & 38 & 0.9868 & 24 & 27 & 60 & 0.9694 & 17 & 36 & 82 & 0.9485 & 13 \\
18 & 39 & 0.9869 & 24 & 28 & 58 & 0.9669 & 16 & 36 & 83 & 0.9485 & 13 \\
18 & 40 & 0.9869 & 24 & 28 & 59 & 0.9670 & 16 & 36 & 84 & 0.9486 & 13 \\
18 & 41 & 0.9869 & 24 & 28 & 60 & 0.9670 & 16 & 36 & 85 & 0.9486 & 13 \\
18 & 42 & 0.9869 & 24 & 28 & 61 & 0.9671 & 16 & 36 & 86 & 0.9487 & 13 \\
19 & 39 & 0.9852 & 23 & 28 & 62 & 0.9671 & 16 & 37 & 83 & 0.9459 & 13 \\
19 & 40 & 0.9852 & 23 & 28 & 63 & 0.9672 & 16 & 37 & 84 & 0.9460 & 13 \\
19 & 41 & 0.9852 & 23 & 28 & 64 & 0.9673 & 16 & 37 & 85 & 0.9460 & 13 \\
20 & 40 & 0.9835 & 22 & 29 & 63 & 0.9648 & 16 & 37 & 86 & 0.9461 & 13 \\
20 & 41 & 0.9835 & 22 & 29 & 64 & 0.9649 & 16 & 37 & 87 & 0.9461 & 13 \\
20 & 42 & 0.9835 & 22 & 29 & 65 & 0.9650 & 16 & 38 & 82 & 0.9433 & 13 \\
20 & 43 & 0.9835 & 22 & 30 & 64 & 0.9625 & 15 & 38 & 83 & 0.9433 & 13 \\
20 & 44 & 0.9835 & 22 & 30 & 65 & 0.9626 & 15 & 38 & 84 & 0.9434 & 13 \\
20 & 45 & 0.9835 & 22 & 30 & 66 & 0.9626 & 15 & 38 & 85 & 0.9434 & 13 \\
\bottomrule
\end{tabular}
\end{table*}
\begin{table*}[p]
\centering
\caption{Calculated Huff factors and $ \kappa_{\urm{max}} $ for each isotope (continued).}
\label{tab:huff_factors_2}
\begin{tabular}{cccc@{\hspace{5em}}cccc@{\hspace{5em}}cccc}
\toprule
Z & A & Huff factor & $\kappa_\mathrm{max}$ & Z & A & Huff factor & $\kappa_\mathrm{max}$ & Z & A & Huff factor & $\kappa_\mathrm{max}$ \\
\midrule
38 & 86 & 0.9435 & 13 & 45 & 103 & 0.9259 & 11 & 51 & 124 & 0.9121 & 10 \\
38 & 87 & 0.9435 & 13 & 45 & 104 & 0.9261 & 11 & 51 & 125 & 0.9123 & 10 \\
38 & 88 & 0.9436 & 13 & 45 & 105 & 0.9264 & 11 & 52 & 120 & 0.9092 & 10 \\
38 & 89 & 0.9436 & 13 & 46 & 100 & 0.9227 & 11 & 52 & 121 & 0.9093 & 10 \\
38 & 90 & 0.9437 & 13 & 46 & 101 & 0.9228 & 11 & 52 & 122 & 0.9094 & 10 \\
39 & 87 & 0.9408 & 12 & 46 & 102 & 0.9229 & 11 & 52 & 123 & 0.9094 & 10 \\
39 & 88 & 0.9409 & 12 & 46 & 103 & 0.9231 & 11 & 52 & 124 & 0.9095 & 10 \\
39 & 89 & 0.9409 & 12 & 46 & 104 & 0.9233 & 11 & 52 & 125 & 0.9096 & 10 \\
39 & 90 & 0.9410 & 12 & 46 & 105 & 0.9235 & 11 & 52 & 126 & 0.9097 & 10 \\
39 & 91 & 0.9411 & 12 & 46 & 106 & 0.9237 & 11 & 52 & 127 & 0.9099 & 10 \\
40 & 88 & 0.9382 & 12 & 46 & 107 & 0.9239 & 11 & 52 & 128 & 0.9100 & 10 \\
40 & 89 & 0.9383 & 12 & 46 & 108 & 0.9241 & 11 & 52 & 129 & 0.9101 & 10 \\
40 & 90 & 0.9383 & 12 & 46 & 109 & 0.9242 & 11 & 52 & 130 & 0.9102 & 10 \\
40 & 91 & 0.9384 & 12 & 46 & 110 & 0.9244 & 11 & 53 & 123 & 0.9070 & 10 \\
40 & 92 & 0.9385 & 12 & 47 & 105 & 0.9206 & 11 & 53 & 124 & 0.9071 & 10 \\
40 & 93 & 0.9386 & 12 & 47 & 106 & 0.9208 & 11 & 53 & 125 & 0.9072 & 10 \\
40 & 94 & 0.9387 & 12 & 47 & 107 & 0.9210 & 11 & 53 & 126 & 0.9072 & 10 \\
40 & 95 & 0.9387 & 12 & 47 & 108 & 0.9212 & 11 & 53 & 127 & 0.9073 & 10 \\
40 & 96 & 0.9388 & 12 & 47 & 109 & 0.9214 & 11 & 53 & 128 & 0.9074 & 10 \\
41 & 91 & 0.9357 & 12 & 47 & 110 & 0.9215 & 11 & 53 & 129 & 0.9075 & 10 \\
41 & 92 & 0.9358 & 12 & 47 & 111 & 0.9217 & 11 & 53 & 130 & 0.9076 & 10 \\
41 & 93 & 0.9359 & 12 & 48 & 106 & 0.9178 & 11 & 53 & 131 & 0.9077 & 10 \\
41 & 94 & 0.9359 & 12 & 48 & 107 & 0.9180 & 11 & 54 & 125 & 0.9047 & 10 \\
41 & 95 & 0.9360 & 12 & 48 & 108 & 0.9183 & 11 & 54 & 126 & 0.9048 & 10 \\
42 & 92 & 0.9330 & 12 & 48 & 109 & 0.9184 & 11 & 54 & 127 & 0.9049 & 10 \\
42 & 93 & 0.9331 & 12 & 48 & 110 & 0.9186 & 11 & 54 & 128 & 0.9049 & 10 \\
42 & 94 & 0.9332 & 12 & 48 & 111 & 0.9188 & 11 & 54 & 129 & 0.9049 & 10 \\
42 & 95 & 0.9333 & 12 & 48 & 112 & 0.9190 & 11 & 54 & 130 & 0.9049 & 10 \\
42 & 96 & 0.9334 & 12 & 48 & 113 & 0.9191 & 11 & 54 & 131 & 0.9051 & 10 \\
42 & 97 & 0.9335 & 12 & 48 & 114 & 0.9192 & 11 & 54 & 132 & 0.9052 & 10 \\
42 & 98 & 0.9336 & 12 & 48 & 115 & 0.9194 & 11 & 54 & 133 & 0.9053 & 10 \\
42 & 99 & 0.9339 & 12 & 48 & 116 & 0.9195 & 11 & 54 & 134 & 0.9055 & 10 \\
42 & 100 & 0.9342 & 12 & 49 & 113 & 0.9162 & 10 & 54 & 135 & 0.9056 & 10 \\
43 & 95 & 0.9305 & 11 & 49 & 114 & 0.9163 & 10 & 54 & 136 & 0.9057 & 10 \\
43 & 96 & 0.9306 & 11 & 49 & 115 & 0.9165 & 10 & 55 & 133 & 0.9027 & 10 \\
43 & 97 & 0.9307 & 11 & 50 & 112 & 0.9131 & 10 & 55 & 134 & 0.9028 & 10 \\
43 & 98 & 0.9308 & 11 & 50 & 113 & 0.9132 & 10 & 55 & 135 & 0.9030 & 10 \\
43 & 99 & 0.9309 & 11 & 50 & 114 & 0.9134 & 10 & 55 & 136 & 0.9031 & 10 \\
44 & 96 & 0.9279 & 11 & 50 & 115 & 0.9135 & 10 & 55 & 137 & 0.9032 & 10 \\
44 & 97 & 0.9280 & 11 & 50 & 116 & 0.9137 & 10 & 56 & 130 & 0.9001 & 10 \\
44 & 98 & 0.9281 & 11 & 50 & 117 & 0.9138 & 10 & 56 & 131 & 0.9001 & 10 \\
44 & 99 & 0.9282 & 11 & 50 & 118 & 0.9140 & 10 & 56 & 132 & 0.9002 & 10 \\
44 & 100 & 0.9283 & 11 & 50 & 119 & 0.9141 & 10 & 56 & 133 & 0.9002 & 10 \\
44 & 101 & 0.9284 & 11 & 50 & 120 & 0.9143 & 10 & 56 & 134 & 0.9002 & 10 \\
44 & 102 & 0.9285 & 11 & 50 & 121 & 0.9144 & 10 & 56 & 135 & 0.9003 & 10 \\
44 & 103 & 0.9287 & 11 & 50 & 122 & 0.9145 & 10 & 56 & 136 & 0.9005 & 10 \\
44 & 104 & 0.9290 & 11 & 50 & 123 & 0.9147 & 10 & 56 & 137 & 0.9006 & 10 \\
44 & 105 & 0.9294 & 11 & 50 & 124 & 0.9148 & 10 & 56 & 138 & 0.9007 & 10 \\
44 & 106 & 0.9298 & 11 & 50 & 125 & 0.9149 & 10 & 57 & 137 & 0.8980 & 9 \\
45 & 99 & 0.9254 & 11 & 50 & 126 & 0.9150 & 10 & 57 & 138 & 0.8981 & 9 \\
45 & 100 & 0.9255 & 11 & 51 & 121 & 0.9118 & 10 & 57 & 139 & 0.8982 & 9 \\
45 & 101 & 0.9256 & 11 & 51 & 122 & 0.9119 & 10 & 58 & 134 & 0.8954 & 9 \\
45 & 102 & 0.9258 & 11 & 51 & 123 & 0.9120 & 10 & 58 & 135 & 0.8954 & 9 \\
\bottomrule
\end{tabular}
\end{table*}
\begin{table*}[p]
\centering
\caption{Calculated Huff factors and $ \kappa_{\urm{max}} $ for each isotope (continued).}
\label{tab:huff_factors_3}
\begin{tabular}{cccc@{\hspace{5em}}cccc@{\hspace{5em}}cccc}
\toprule
Z & A & Huff factor & $\kappa_\mathrm{max}$ & Z & A & Huff factor & $\kappa_\mathrm{max}$ & Z & A & Huff factor & $\kappa_\mathrm{max}$ \\
\midrule
58 & 136 & 0.8953 & 9 & 66 & 156 & 0.8791 & 9 & 72 & 177 & 0.8684 & 8 \\
58 & 137 & 0.8954 & 9 & 66 & 157 & 0.8794 & 9 & 72 & 178 & 0.8686 & 8 \\
58 & 138 & 0.8955 & 9 & 66 & 158 & 0.8797 & 9 & 72 & 179 & 0.8687 & 8 \\
58 & 139 & 0.8956 & 9 & 66 & 159 & 0.8799 & 9 & 72 & 180 & 0.8689 & 8 \\
58 & 140 & 0.8957 & 9 & 66 & 160 & 0.8802 & 9 & 72 & 181 & 0.8691 & 8 \\
58 & 141 & 0.8959 & 9 & 66 & 161 & 0.8804 & 9 & 72 & 182 & 0.8693 & 8 \\
58 & 142 & 0.8960 & 9 & 66 & 162 & 0.8807 & 9 & 73 & 177 & 0.8659 & 8 \\
58 & 143 & 0.8963 & 9 & 66 & 163 & 0.8809 & 9 & 73 & 178 & 0.8660 & 8 \\
58 & 144 & 0.8966 & 9 & 66 & 164 & 0.8811 & 9 & 73 & 179 & 0.8661 & 8 \\
59 & 141 & 0.8932 & 9 & 67 & 163 & 0.8783 & 9 & 73 & 180 & 0.8663 & 8 \\
59 & 142 & 0.8934 & 9 & 67 & 164 & 0.8785 & 9 & 73 & 181 & 0.8665 & 8 \\
59 & 143 & 0.8935 & 9 & 67 & 165 & 0.8787 & 9 & 73 & 182 & 0.8666 & 8 \\
60 & 142 & 0.8907 & 9 & 67 & 166 & 0.8789 & 9 & 73 & 183 & 0.8668 & 8 \\
60 & 143 & 0.8909 & 9 & 67 & 167 & 0.8792 & 9 & 74 & 180 & 0.8636 & 8 \\
60 & 144 & 0.8910 & 9 & 68 & 160 & 0.8748 & 8 & 74 & 181 & 0.8638 & 8 \\
60 & 145 & 0.8913 & 9 & 68 & 161 & 0.8751 & 8 & 74 & 182 & 0.8640 & 8 \\
60 & 146 & 0.8915 & 9 & 68 & 162 & 0.8754 & 8 & 74 & 183 & 0.8642 & 8 \\
60 & 147 & 0.8919 & 9 & 68 & 163 & 0.8757 & 8 & 74 & 184 & 0.8644 & 8 \\
60 & 148 & 0.8924 & 9 & 68 & 164 & 0.8759 & 8 & 74 & 185 & 0.8645 & 8 \\
60 & 149 & 0.8930 & 9 & 68 & 165 & 0.8761 & 8 & 74 & 186 & 0.8647 & 8 \\
60 & 150 & 0.8936 & 9 & 68 & 166 & 0.8763 & 8 & 75 & 185 & 0.8619 & 8 \\
61 & 145 & 0.8886 & 9 & 68 & 167 & 0.8766 & 8 & 75 & 186 & 0.8620 & 8 \\
61 & 146 & 0.8888 & 9 & 68 & 168 & 0.8768 & 8 & 75 & 187 & 0.8622 & 8 \\
61 & 147 & 0.8891 & 9 & 68 & 169 & 0.8770 & 8 & 76 & 184 & 0.8591 & 8 \\
62 & 144 & 0.8858 & 9 & 68 & 170 & 0.8772 & 8 & 76 & 185 & 0.8592 & 8 \\
62 & 145 & 0.8860 & 9 & 68 & 171 & 0.8774 & 8 & 76 & 186 & 0.8594 & 8 \\
62 & 146 & 0.8861 & 9 & 68 & 172 & 0.8776 & 8 & 76 & 187 & 0.8595 & 8 \\
62 & 147 & 0.8864 & 9 & 69 & 167 & 0.8740 & 8 & 76 & 188 & 0.8597 & 8 \\
62 & 148 & 0.8867 & 9 & 69 & 168 & 0.8742 & 8 & 76 & 189 & 0.8599 & 8 \\
62 & 149 & 0.8872 & 9 & 69 & 169 & 0.8745 & 8 & 76 & 190 & 0.8602 & 8 \\
62 & 150 & 0.8877 & 9 & 69 & 170 & 0.8747 & 8 & 76 & 191 & 0.8602 & 8 \\
62 & 151 & 0.8882 & 9 & 69 & 171 & 0.8749 & 8 & 76 & 192 & 0.8603 & 8 \\
62 & 152 & 0.8888 & 9 & 70 & 166 & 0.8712 & 8 & 76 & 193 & 0.8603 & 8 \\
62 & 153 & 0.8890 & 9 & 70 & 167 & 0.8714 & 8 & 76 & 194 & 0.8603 & 8 \\
62 & 154 & 0.8893 & 9 & 70 & 168 & 0.8717 & 8 & 77 & 188 & 0.8570 & 8 \\
63 & 150 & 0.8848 & 9 & 70 & 169 & 0.8719 & 8 & 77 & 189 & 0.8572 & 8 \\
63 & 151 & 0.8853 & 9 & 70 & 170 & 0.8722 & 8 & 77 & 190 & 0.8574 & 8 \\
63 & 152 & 0.8858 & 9 & 70 & 171 & 0.8724 & 8 & 77 & 191 & 0.8576 & 8 \\
63 & 153 & 0.8863 & 9 & 70 & 172 & 0.8726 & 8 & 77 & 192 & 0.8577 & 8 \\
64 & 152 & 0.8829 & 9 & 70 & 173 & 0.8728 & 8 & 77 & 193 & 0.8578 & 8 \\
64 & 153 & 0.8834 & 9 & 70 & 174 & 0.8730 & 8 & 77 & 194 & 0.8579 & 8 \\
64 & 154 & 0.8839 & 9 & 70 & 175 & 0.8731 & 8 & 78 & 190 & 0.8547 & 8 \\
64 & 155 & 0.8842 & 9 & 70 & 176 & 0.8733 & 8 & 78 & 191 & 0.8549 & 8 \\
64 & 156 & 0.8845 & 9 & 70 & 177 & 0.8735 & 8 & 78 & 192 & 0.8551 & 8 \\
64 & 157 & 0.8848 & 9 & 71 & 173 & 0.8703 & 8 & 78 & 193 & 0.8552 & 8 \\
64 & 158 & 0.8850 & 9 & 71 & 174 & 0.8705 & 8 & 78 & 194 & 0.8553 & 8 \\
64 & 159 & 0.8852 & 9 & 71 & 175 & 0.8706 & 8 & 78 & 195 & 0.8555 & 8 \\
64 & 160 & 0.8855 & 9 & 71 & 176 & 0.8708 & 8 & 78 & 196 & 0.8556 & 8 \\
65 & 157 & 0.8821 & 9 & 72 & 172 & 0.8676 & 8 & 78 & 197 & 0.8558 & 8 \\
65 & 158 & 0.8823 & 9 & 72 & 173 & 0.8678 & 8 & 78 & 198 & 0.8559 & 8 \\
65 & 159 & 0.8826 & 9 & 72 & 174 & 0.8680 & 8 & 79 & 195 & 0.8530 & 8 \\
66 & 154 & 0.8782 & 9 & 72 & 175 & 0.8681 & 8 & 79 & 196 & 0.8532 & 8 \\
66 & 155 & 0.8786 & 9 & 72 & 176 & 0.8683 & 8 & 79 & 197 & 0.8533 & 8 \\
\bottomrule
\end{tabular}
\end{table*}
\begin{table*}[t]
\centering
\caption{Calculated Huff factors and $ \kappa_{\urm{max}} $ for each isotope (continued).}
\label{tab:huff_factors_4}
\begin{tabular}{cccc@{\hspace{5em}}cccc@{\hspace{5em}}cccc}
\toprule
Z & A & Huff factor & $\kappa_\mathrm{max}$ & Z & A & Huff factor & $\kappa_\mathrm{max}$ & Z & A & Huff factor & $\kappa_\mathrm{max}$ \\
\midrule
79 & 198 & 0.8534 & 8 & 84 & 208 & 0.8431 & 7 & 90 & 232 & 0.8370 & 7 \\
79 & 199 & 0.8535 & 8 & 84 & 209 & 0.8433 & 7 & 90 & 233 & 0.8373 & 7 \\
80 & 194 & 0.8503 & 8 & 84 & 210 & 0.8434 & 7 & 90 & 234 & 0.8376 & 7 \\
80 & 195 & 0.8505 & 8 & 85 & 210 & 0.8413 & 7 & 91 & 229 & 0.8341 & 7 \\
80 & 196 & 0.8507 & 8 & 85 & 211 & 0.8415 & 7 & 91 & 230 & 0.8344 & 7 \\
80 & 197 & 0.8509 & 8 & 86 & 219 & 0.8410 & 7 & 91 & 231 & 0.8347 & 7 \\
80 & 198 & 0.8510 & 8 & 86 & 220 & 0.8412 & 7 & 91 & 232 & 0.8349 & 7 \\
80 & 199 & 0.8510 & 8 & 86 & 221 & 0.8420 & 7 & 91 & 233 & 0.8352 & 7 \\
80 & 200 & 0.8510 & 8 & 86 & 222 & 0.8428 & 7 & 91 & 234 & 0.8355 & 7 \\
80 & 201 & 0.8512 & 8 & 86 & 223 & 0.8430 & 7 & 92 & 233 & 0.8331 & 7 \\
80 & 202 & 0.8514 & 8 & 86 & 224 & 0.8431 & 7 & 92 & 234 & 0.8333 & 7 \\
80 & 203 & 0.8515 & 8 & 87 & 221 & 0.8397 & 7 & 92 & 235 & 0.8336 & 7 \\
80 & 204 & 0.8517 & 8 & 87 & 222 & 0.8402 & 7 & 92 & 236 & 0.8339 & 7 \\
81 & 203 & 0.8490 & 8 & 87 & 223 & 0.8408 & 7 & 92 & 237 & 0.8341 & 7 \\
81 & 204 & 0.8492 & 8 & 88 & 223 & 0.8385 & 7 & 92 & 238 & 0.8344 & 7 \\
81 & 205 & 0.8494 & 8 & 88 & 224 & 0.8388 & 7 & 93 & 235 & 0.8315 & 7 \\
82 & 204 & 0.8467 & 8 & 88 & 225 & 0.8392 & 7 & 93 & 236 & 0.8318 & 7 \\
82 & 205 & 0.8469 & 8 & 88 & 226 & 0.8395 & 7 & 93 & 237 & 0.8321 & 7 \\
82 & 206 & 0.8471 & 8 & 88 & 227 & 0.8398 & 7 & 93 & 238 & 0.8323 & 7 \\
82 & 207 & 0.8472 & 8 & 88 & 228 & 0.8402 & 7 & 93 & 239 & 0.8326 & 7 \\
82 & 208 & 0.8474 & 8 & 89 & 225 & 0.8370 & 7 & 94 & 238 & 0.8303 & 7 \\
82 & 209 & 0.8476 & 8 & 89 & 226 & 0.8374 & 7 & 94 & 239 & 0.8305 & 7 \\
82 & 210 & 0.8478 & 8 & 89 & 227 & 0.8377 & 7 & 94 & 240 & 0.8308 & 7 \\
83 & 207 & 0.8451 & 7 & 90 & 228 & 0.8359 & 7 & 94 & 241 & 0.8310 & 7 \\
83 & 208 & 0.8453 & 7 & 90 & 229 & 0.8362 & 7 & 94 & 242 & 0.8313 & 7 \\
83 & 209 & 0.8454 & 7 & 90 & 230 & 0.8365 & 7 & 94 & 243 & 0.8315 & 7 \\
83 & 210 & 0.8456 & 7 & 90 & 231 & 0.8368 & 7 & 94 & 244 & 0.8317 & 7 \\
\bottomrule
\end{tabular}
\end{table*}

\par
Tables~\ref{tab:huff_factors_1}--\ref{tab:huff_factors_4} present the calculated Huff factors and $ \kappa_{\urm{max}} $ for each isotope used in the present study.
\section{Convergence of the partial wave expansion}
\label{app:kappa_convergence}
\par
We discuss the convergence properties of the partial wave expansion appearing in Eq.~\eqref{eq:DIO_spectrum_kappa_sum}.
The index $ \kappa $ is an integer running over $ - \infty < \kappa < \infty $, excluding zero.
In practice, the dominant contributions come from terms with small $ \left| \kappa \right| $,
because the overlap between the bound muon and the components of the electron wave function with large $ \left| \kappa \right| $ is suppressed.
Therefore, the summation can be truncated with a sufficiently small truncation error in numerical calculations.
\par
We define the contribution from each $ \kappa $ as
\begin{equation}
  Q_{\kappa}
  =
  \frac{1}{\Gamma_0}
  \int_{m_e}^{m_{\mu} - B_{\mu}}
  \frac{d \Gamma_{\kappa}}{d E_e}
  \, dE_e,
\end{equation}
so that
\begin{equation}
  Q
  =
  \sum_{\kappa}
  Q_{\kappa}.
\end{equation}
Note that $ Q_{-\kappa} \simeq Q_{\kappa} $;
if the electron mass is neglected, the equality holds exactly.
Therefore, it suffices to focus on the absolute value of $ \kappa $ when discussing the convergence.
\par
\begin{figure}[tb]
  \centering
  \includegraphics[width=1.0\linewidth]{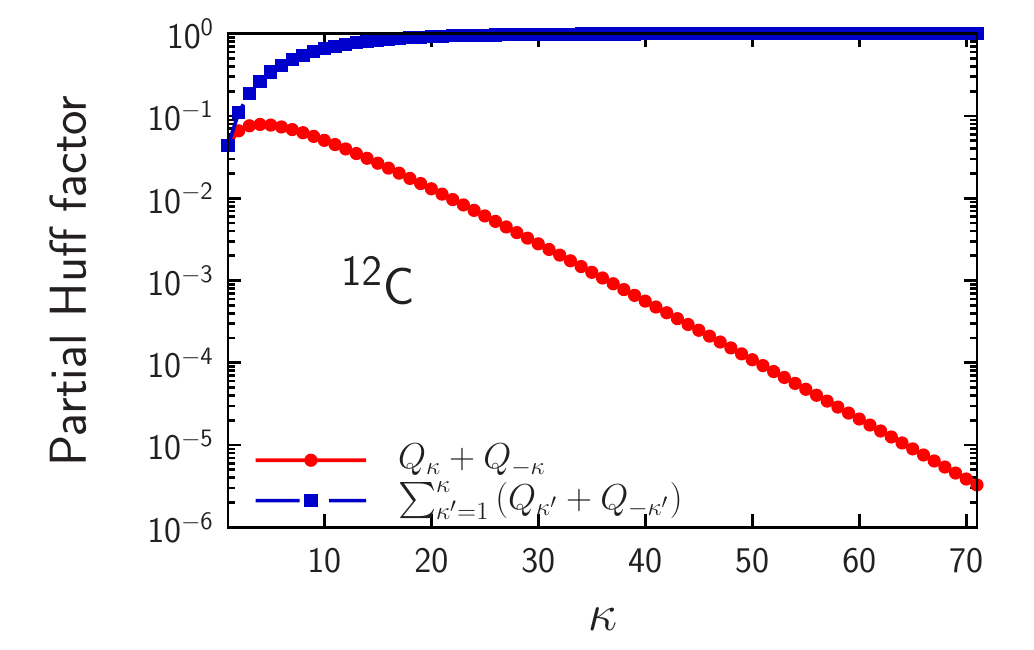}
  \includegraphics[width=1.0\linewidth]{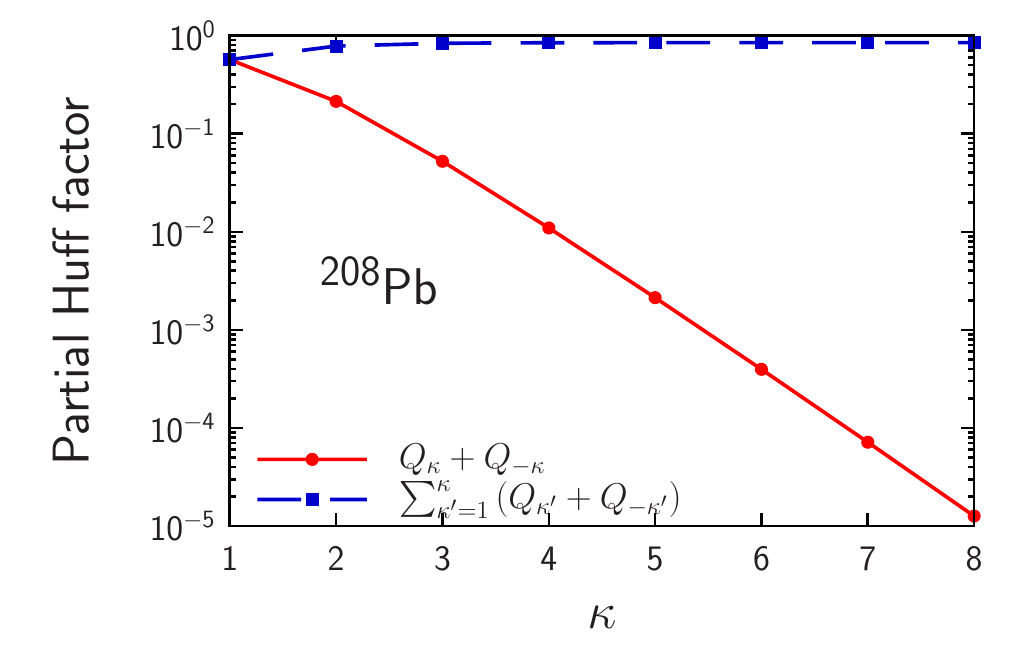}
  \caption{Contribution of different $ \kappa $ for the Huff factor.
    The top and bottom panels are for $ \nuc{C}{12}{} $ and $ \nuc{Pb}{208}{} $, respectively. 
    The red circles are the $ \left| \kappa \right| $ contribution to the Huff factor,
    $ Q_{\kappa} + Q_{-\kappa} $,
    and the blue squares are its cumulative sum $ \sum_{\kappa' = 1}^{\kappa} \left( Q_{\kappa'} + Q_{-\kappa'} \right)$.
    Lines are guide for the eye.}
  \label{fig:kappa_conv}
\end{figure}

The contributions of each $ \kappa $ are shown in Fig.~\ref{fig:kappa_conv} for $ \nuc{C}{12}{} $ and $ \nuc{Pb}{208}{} $.
The horizontal axis represents the absolute value of $ \kappa $.
The red circles indicate $ Q_{\kappa} + Q_{-\kappa} $,
while the blue squares show the cumulative sum
$ \sum_{\kappa' = 1}^{\kappa} \left( Q_{\kappa'} + Q_{-\kappa'} \right) $.
\par
As expected, the contributions decrease with increasing $ \left| \kappa \right| $, thus justifying the truncation.
For heavier nuclei, the wave function of the bound muon is more localized near the nuclear center, which suppresses transitions to electron states with large angular momentum.
Consequently, the sum over $ \kappa $ converges more rapidly for heavier nuclei.
To achieve a truncation error below $ 0.1 \, \% $,
contributions up to $ \left| \kappa \right| \leq 5 $ are required for $ \nuc{Pb}{208}{} $.
In contrast, the convergence is slower for lighter nuclei;
contributions up to $ \left| \kappa \right| \leq 46 $ are necessary for $ \nuc{C}{12}{} $. 
\section{Distortion effects of the emitted electron compared to the plane wave approximation}
\label{app:distortion}
\par
The calculation of the Huff factor depends on the treatment of the lepton wave functions.
In this appendix, we first derive a simplified formula obtained by neglecting the Coulomb distortion of the emitted electron, namely under the plane wave approximation.
We then use this simplified framework to compare, in a step-by-step manner, the effects of the electron distortion, the relativistic treatment of the bound muon, and the finite-size effect of the nucleus on the Huff factor.
\par
In the body of this article, we employed the partial wave expansion in order to account for the Coulomb distortion of the emitted electron.
In contrast, treating the electron as a plane wave is a crude approximation, but it leads to a simpler formula than Eq.~\eqref{eq:DIO_spectrum}.
Here, we present the decay-rate formula neglecting the Coulomb distortion of the electron.
Since the partial wave expansion is not used in this formalism, truncation errors do not arise.
\par
If the electron wave function is approximated by a plane wave, the Huff factor is given by
\begin{align}
  Q
  & = 
    \frac{16}{\pi m_{\mu}^5 f(m_e/m_\mu)}
    \int_{m_e}^{m_{\mu} - B_{\mu}}
    d E_e
    \, p
    \int_0^{K_0}
    dK
    \, K^2
    \int_{-1}^{1}
    dx
    \notag \\    
  & \quad
    \times
    \left[
    \left(
    \left| \tilde{g}_{\mu} \left( q \right) \right|^2
    +
    \left| \tilde{f}_{\mu} \left( q \right) \right|^2
    \right)
    \left(
    3 E_e K_0^2
    -
    E_e K^2
    -
    2 p K_0 K x
    \right)
    \right.
    \notag \\
  & \qquad
    +
    \frac{2}{q}
    \tilde{g}_{\mu} \left( q \right)
    \tilde{f}_{\mu} \left( q \right)
    \left\{ p^2 \left( K_0^2 - K^2 \right)
    +
    2 E_e K_0 K^2
    \right. \
    \notag \\
  & \qquad
    \left.
    \left.
    +
    \left( 2 E_e K_0 + K_0^2 - 3K^2 \right)
    p K x
    -
    2 p^2 K^2 x^2
    \right\}
    \right],
    \label{eq:huff_factor_pwa}
\end{align}
where $ p = \sqrt{E_e^2 - m_e^2} $ and $ q = \sqrt{p^2 + K^2 + 2pKx} $.
Here, $ \tilde{g}_{\mu} \left( q \right) $ and $ \tilde{f}_{\mu} \left( q \right) $ are defined as
\begin{align}
  \tilde{g}_{\mu} \left( q \right)
  & = 
    \int_0^{\infty}
    j_0 \left( qr \right)
    g_{\mu} \left( r \right)
    r^2 \, dr, \\
  \tilde{f}_{\mu} \left( q \right)
  & = 
    \int_0^{\infty}
    j_1 \left( qr \right)
    f_{\mu} \left( r \right)
    r^2 \, dr.
\end{align}
These radial wave functions of the muon, $ g_{\mu} \left( r \right) $ and $ f_{\mu} \left( r \right) $, are numerically obtained by solving Eq.~\eqref{eq:Dirac_eq}.
\par
Using this formula, we now discuss the behavior of the Huff factor for small $\zeta = Z\alpha$ under the plane wave approximation for the electron.
Here, $\alpha\simeq 1/137$ is the fine-structure constant.
Neglecting the electron mass, the integrals over $E_e$ and $K$ can be carried out analytically, yielding
\begin{align}
  Q
  & =
    \frac{2}{\pi m_{\mu}^3}
    \int_0^{m_{\mu} - B_{\mu}}
    dq \,
    q^2
    \left\{
    \left(
    m_{\mu} - B_{\mu}
    \right)^2
    -
    q^2
    \right\}
    \notag \\
  & \quad
    \times
    \left[
    \left(
    m_{\mu} - B_{\mu}
    \right)
    \left(
    \left| \tilde{g}_{\mu} \left( q \right) \right|^2
    +
    \left| \tilde{f}_{\mu} \left( q \right) \right|^2
    \right)
    +
    2q\tilde{g}_{\mu} \left( q \right)
    \tilde{f}_{\mu} \left( q \right)\right],
\end{align}
for general muon radial wave functions.
We note that the integral variable $ q $ is interpreted as the magnitude of the intrinsic momentum of the bound muon.
\par
We now consider the solution of the Dirac equation with a point-charge Coulomb potential,
\begin{align}
  g_{\mu} \left( r \right)
  & = 
    2
    \left( m_{\mu} \zeta \right)^{3/2}
    \sqrt{\frac{1+\gamma}{\Gamma\left(1+2\gamma\right)}}\left(2m_\mu\zeta r\right)^{\gamma-1}\exp \left( - m_{\mu} \zeta r \right),
    \label{eq:rela_wave_g} \\
  f_{\mu} \left( r \right)
  & = 
    -2
    \left( m_{\mu} \zeta \right)^{3/2}
    \sqrt{\frac{1-\gamma}{\Gamma\left(1+2\gamma\right)}}\left(2m_\mu\zeta r\right)^{\gamma-1}\exp \left( - m_{\mu} \zeta r \right)
    \label{eq:rela_wave_f}
\end{align}
with the binding energy $ B_{\mu} = m_{\mu} \left(1-\gamma\right) $ and $\gamma=\sqrt{1-\zeta^2}$.
In this case, the Huff factor reduces to
\begin{align}
  Q
  & =
    \frac{2^{2\gamma+1} \gamma \Gamma \left( \gamma \right)^2}{\pi \Gamma \left( 1 + 2\gamma \right)}
    \int_0^{\arcsin \gamma}
    d \theta
    \,
    \left( \gamma^2 - \sin^2 \theta \right)^2
    \sin^2\theta
    \cos^{2\gamma-4}\theta
    \notag \\
  & \quad
    \times
    \left[
    \gamma^2
    \left( 1 + \gamma \right)
    U_{\gamma} \left( \theta \right)^2
    +
    \left( 1 - \gamma \right)
    U'_{\gamma} \left( \theta \right)^2
    \right.
    \notag \\
  & \qquad
    \left.
    +
    2 \left( 1 - \gamma^2 \right)
    U_{\gamma} \left( \theta \right)
    U'_{\gamma} \left( \theta \right)
    \tan \theta
    \right],
    \label{eq:analytic_Huff_factor_rela_muon}
\end{align}
where
\begin{equation}
  U_{\gamma} \left( \theta \right)
  =
  \frac{\sin \left\{ \left( 1 + \gamma \right) \theta \right\}}{\sin\theta}
\end{equation}
and $ U'_{\gamma} \left( \theta \right) = \partial U_{\gamma} \left( \theta \right)/\partial\theta$.
Note that $ Q $ depends on $ \zeta $ only through $ \gamma = \sqrt{1 - \zeta^2}$ in Eq.~\eqref{eq:analytic_Huff_factor_rela_muon}.
We find analytically that
\begin{equation}
  \left.
    \frac{dQ}{d\gamma}
  \right|_{\gamma=1}
  =
  11,
\end{equation}
and therefore the expansion around $\gamma=1$ is given by
\begin{equation}
  Q
  =
  1
  +
  11
  \left( \gamma - 1 \right)
  +
  \mathcal{O} \left(\left(\gamma-1\right)^2\right),
\end{equation}
which leads to
\begin{equation}
  \label{eq:analytic_rela_up_to_zeta_squared}
  Q
  =
  1
  -
  \frac{11}{2}\zeta^2
  +
  \mathcal{O}\left(\zeta^4\right).
\end{equation}
The behavior of the Huff factor at small $\zeta$ is consistent with the plane-wave analyses in Refs.~\cite{Huff1961-ur,Uberall:1960zz}.
\par
As a further simplification, we consider the nonrelativistic muon wave function obtained from the Schr\"{o}dinger equation with a point-charge Coulomb potential,
\begin{align}
  g_{\mu} \left( r \right)
  & = 
    2
    \left( m_{\mu} \zeta \right)^{3/2}
    \exp \left( - m_{\mu} \zeta r \right),
    \label{eq:nonrela_wave_g} \\
  f_{\mu} \left( r \right)
  & = 
    0,
    \label{eq:nonrela_wave_f}
\end{align}
with the binding energy $ B_{\mu} = m_{\mu} \zeta^2 / 2 $.
Then we find the analytic result,
\begin{align}
  Q
  & = 
    \frac{2 - \zeta^2}{48 \pi \left( 4 + \zeta^4 \right)}
    \left\{
    3
    \left(
    64
    -
    256 \zeta^2
    +
    560 \zeta^4
    -
    128 \zeta^6
    \right.
    \right.
    \notag \\
  & \quad
    \left.
    \left.
    +
    140 \zeta^8
    -
    16 \zeta^{10}
    +
    \zeta^{12}
    \right)
    \arctan \left( \frac{2 - \zeta^2}{2 \zeta} \right)
    \right.
    \notag \\
  & \quad
    \left.
    +
    2 \zeta
    \left(
    96
    -
    368 \zeta^2
    -
    48\zeta^4
    +
    24 \zeta^6
    +
    46\zeta^8
    -
    3\zeta^{10}
    \right)
    \right\}
    \label{eq:analytic_Huff_factor_nonrela_muon} \\
  & = 
    1
    -
    \frac{9}{2}
    \zeta^2
    +
    \mathcal{O} \left( \zeta^4 \right).
    \label{eq:analytic_nonrela_up_to_zeta_squared}
\end{align}
An expression equivalent to Eq.~\eqref{eq:analytic_Huff_factor_nonrela_muon} was given in Ref.~\cite{PhysRev.120.1450}.
We remark that, since Eqs.~\eqref{eq:nonrela_wave_g} and \eqref{eq:nonrela_wave_f} represent only the leading-order terms in the $\gamma \to 1$ limit of Eqs.~\eqref{eq:rela_wave_g} and \eqref{eq:rela_wave_f}, the coefficients of the $\zeta^2$ and higher-order terms differ between Eqs.~\eqref{eq:analytic_rela_up_to_zeta_squared} and \eqref{eq:analytic_nonrela_up_to_zeta_squared}.
We also note that Eq.~\eqref{eq:analytic_nonrela_up_to_zeta_squared} yields a larger Huff factor than Eq.~\eqref{eq:analytic_rela_up_to_zeta_squared} for any relevant $\zeta$.
\par
\begin{figure}[tb]
  \centering
  \includegraphics[width=1.0\linewidth]{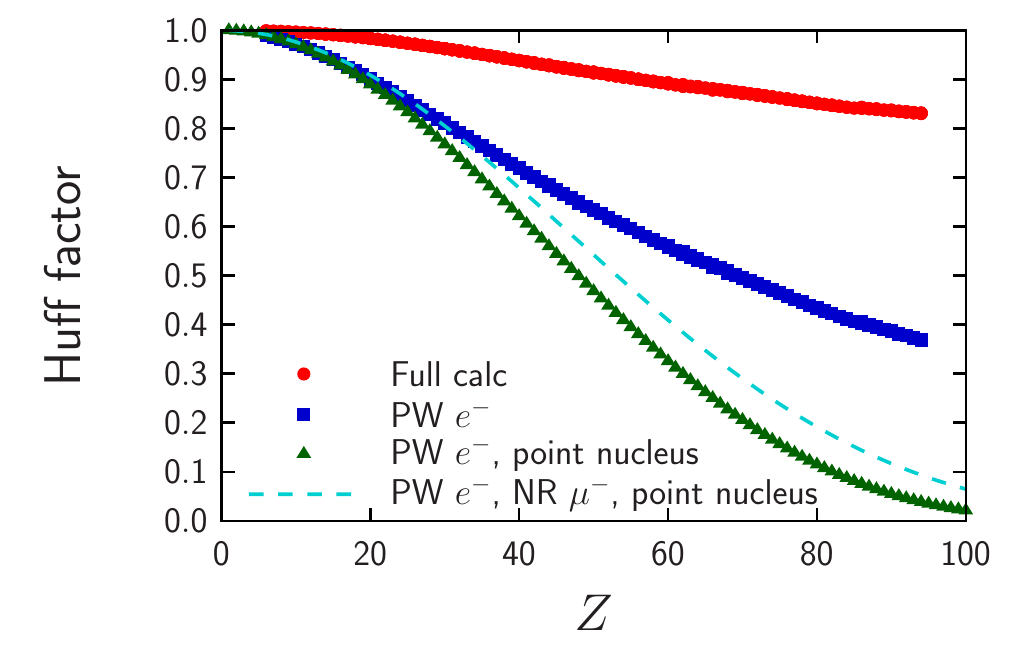}
  \caption{Comparison of the Huff factor obtained with different treatments of the lepton wave functions.
    The red circles show the full calculation including the Coulomb distortion of the emitted electron, while the blue squares show the corresponding results in the plane wave approximation for the electron.
    The green triangles correspond to the case in which the electron is treated as a plane wave and the bound muon is described by the analytic Dirac solution for a point nucleus, as evaluated from Eq.~\eqref{eq:analytic_Huff_factor_rela_muon}.
    The light-blue dashed curve is given by the analytic formula in Eq.~\eqref{eq:analytic_Huff_factor_nonrela_muon},
    in which the electron wave function is approximated by a plane wave and the bound-muon wave function is obtained from the Schr\"{o}dinger equation with a point-charge Coulomb potential.
    The horizontal axis shows the atomic number $ Z $ of the nucleus
    and isotopes with the same $ Z $ are plotted together.}
  \label{fig:huff_comparison}
\end{figure}

In Fig.~\ref{fig:huff_comparison}, we compare the atomic-number dependence of the Huff factor for several different treatments of the lepton wave functions.
The three series without the Coulomb distortion of the emitted electron correspond to plane-wave results for different treatments of the bound muon and the nucleus.
The blue squares are obtained from Eq.~\eqref{eq:huff_factor_pwa} using the numerical solution of Eq.~\eqref{eq:Dirac_eq}.
The green triangles are obtained by numerically integrating Eq.~\eqref{eq:analytic_Huff_factor_rela_muon} for a point nucleus.
The light-blue dashed line is obtained from the analytic formula in Eq.~\eqref{eq:analytic_Huff_factor_nonrela_muon} for a nonrelativistic bound muon.
The red circles show our full calculation, in which the Coulomb distortion of the emitted electron is included.
\par
The comparison between the full calculation and the plane wave results in Fig.~\ref{fig:huff_comparison} shows that the Coulomb distortion of the emitted electron enhances the total DIO rate, and the enhancement becomes more pronounced for heavier nuclei.
As a result, the $ Z $-dependence of the Huff factor in the full calculation is much milder than that in the plane wave approximation.
The corresponding enhancement factors are $ 1.01 $ for $ \nuc{C}{12}{} $,
$ 1.09 $ for $ \nuc{Ca}{40}{} $,
and $ 2.00 $ for $ \nuc{Pb}{208}{} $.
Therefore, the electron distortion must be taken into account in order to achieve percent-level accuracy over the entire range of $ Z $.
\par
\begin{figure}[tb]
  \centering
  \includegraphics[width=1.0\linewidth]{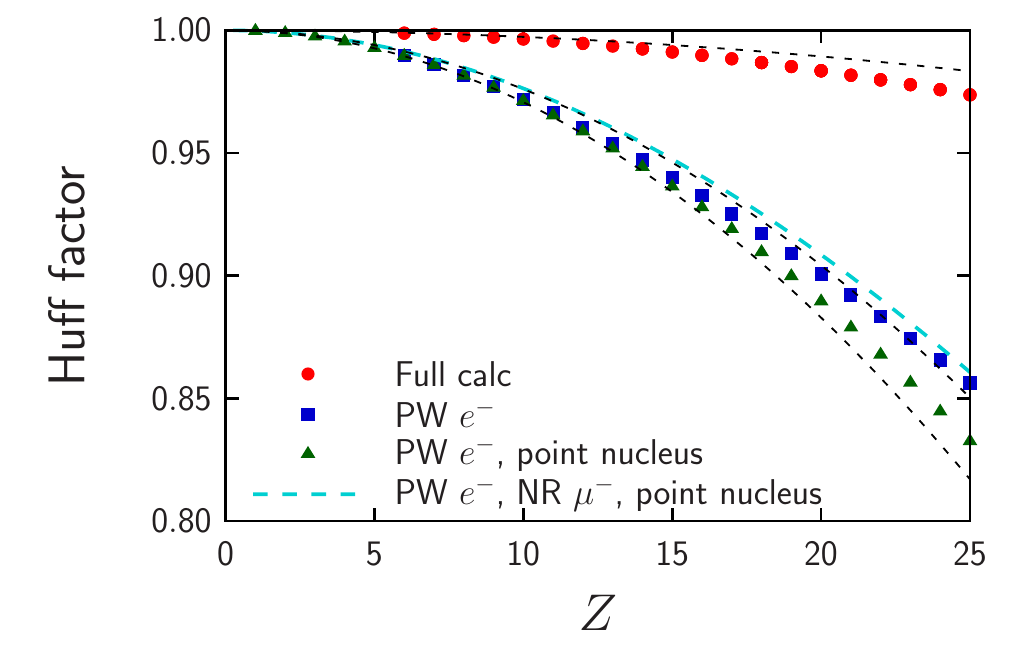}
  \caption{Enlarged view of Fig.~\ref{fig:huff_comparison} for the range $0\le Z\le 25$.
  The three black dashed curves, $Q=1-\zeta^2/2$, $Q=1-9\zeta^2/2$, and $Q=1-11\zeta^2/2$, are also shown for comparison.}
  \label{fig:huff_comparison_small_Z}
\end{figure}

Finally, we discuss the small-$\zeta$ behavior of the Huff factor for the various setups considered above.
Figure~\ref{fig:huff_comparison_small_Z} shows an enlarged view of Fig.~\ref{fig:huff_comparison} for the small-$Z$ region, where a small-$\zeta$ expansion provides a useful description.
For reference, the black dashed curves in the figure represent $Q=1-\zeta^2/2$, $Q=1-9\zeta^2/2$, and $Q=1-11\zeta^2/2$.
\par
Table~\ref{tab:zeta_expansion} compares the Huff factor for $Z=6$ in the same four setups shown in Figs.~\ref{fig:huff_comparison} and \ref{fig:huff_comparison_small_Z}, showing the differences more explicitly.
The first three columns specify the treatments of the electron wave function, the muon wave function, and the nucleus, respectively.
The fourth column lists the Huff factor $Q$ for $Z=6$, and the fifth column lists $ \left( 1 - Q \right)/\zeta^2$, which corresponds to the coefficient $c$ in the small-$\zeta$ form $Q = 1 - c\zeta^2$.
Here, isotope dependence is omitted because it is negligible for such a light nucleus.
\begin{table}[t]
  \caption{
    Comparison of the Huff factor for $Z=6$ under various assumptions for the lepton wave functions.
    The first column indicates whether the electron wave function is treated as a distorted wave (DW) or a plane wave (PW).
    The second column indicates whether the bound-muon wave function is treated relativistically (R) or nonrelativistically (NR).
    The third column indicates whether the lepton wave functions are evaluated for a finite-size nucleus (Finite) or a point-like nucleus (Point).
    The fourth column lists the Huff factor $Q$ for $Z=6$, and the fifth column lists $ \left( 1 - Q \right) / \zeta^2$, which corresponds to the coefficient of $\zeta^2$ in the small-$\zeta$ expansion of $Q$.
    The first row corresponds to our full calculation based on Eq.~\eqref{eq:DIO_spectrum}.}
  \label{tab:zeta_expansion}
  \centering
  \begin{tabular}{lllD{.}{.}{4}D{.}{.}{2}}
    \toprule
    $ e^{-} $ & $ \mu^{-} $ & Nucleus & \multicolumn{1}{c}{$ Q_{Z=6} $} & \multicolumn{1}{c}{$ \left( 1 - Q_{Z=6} \right)/\zeta^2$} \\
    \midrule
    DW & R & Finite & 0.9989 & 0.60 \\
    PW & R & Finite & 0.9896 & 5.42 \\
    PW & R & Point & 0.9895 & 5.47 \\
    PW & NR & Point & 0.9914 & 4.48 \\
    \bottomrule
  \end{tabular}
\end{table}
\par
Equations~\eqref{eq:analytic_nonrela_up_to_zeta_squared} and \eqref{eq:analytic_rela_up_to_zeta_squared} show that the dashed curve and the green triangles behave as $1 - 9\zeta^2/2$ and $1 - 11\zeta^2/2$, respectively, for small $Z$.
Figure~\ref{fig:huff_comparison_small_Z} and the values in Table~\ref{tab:zeta_expansion} indicate that the blue squares are well described by the $1 - 11\zeta^2/2$ behavior in the small-$Z$ region.
This agreement suggests that nuclear finite-size effects are less important for light nuclei.
For heavier nuclei, however, finite-size effects become important, and the result for a finite-size nucleus exceeds even the point-nucleus result with a nonrelativistic muon, as seen in Figs.~\ref{fig:huff_comparison} and \ref{fig:huff_comparison_small_Z}.
\par
As discussed in early studies~\cite{Huff1961-ur,Uberall:1960zz}, once the enhancement due to the Coulomb distortion of the emitted electron is taken into account, the small-$\zeta$ behavior of the Huff factor for a point nucleus was argued to be $Q \simeq 1 - \zeta^2/2$.
This behavior was also numerically supported in a recent study of low-$Z$ nuclei~\cite{Czarnecki:2025phw}.
Our full calculation at small $Z$ is also consistent with the behavior of $Q\simeq 1-\zeta^2/2$.
\bibliographystyle{elsarticle-num-names}
\bibliography{ref}
\end{document}